# How unique are hallucinated citations offered by generative Artificial Intelligence models?

*Dirk H.R. Spennemann*

Gulbali Institute; Charles Sturt University; Albury NSW 2640, Australia
e-mail: dspennemann@csu.edu.au
ORCID: 0000-0003-2639-7950

## Abstract

This paper investigates how generative AI produces and propagates hallucinated academic references, focusing on the recurring non-existent citation "Education Governance and Datafication" attributed to Ben Williamson and Nelli Piattoeva. Drawing on 137 accessible source papers identified through Google Scholar and Google searches, the study analyses the structure, recurrence, and onward citation of this phantom reference. It shows that hallucinated citations are not random inventions but patterned recombinations of real authors, journals, dates, and keywords, with duplication occurring in nearly 30% of cases. The paper also reports a structured interrogation of ChatGPT 5-mini about how it generates citations and finds that, absent verification, the model reconstructs plausible references from learned patterns rather than factual recall. Finally, ten AI-generated essays on datafication and school governance were examined: while most references were genuine or partly accurate, 9.2% remained hallucinated, including an exact match to the most common phantom citation. The findings highlight ongoing risks to academic integrity and show that web-enabled AI still does not fully eliminate fabricated references.

## Keywords

Academic integrity; AI hallucination; generative Artificial Intelligence; fabricated references; ghost citations

## Introduction

Hallucinations in generative Artificial Intelligence (genAI) models are a widely recognized problem. One of the most noticeable forms is the inclusion of fabricated academic references that are added to essays and papers that are written *in toto* or in part by genAI tools ('ghost references,' 'zombie references,' 'phantom references,' 'hallucitations'). Although the overall occurrence of hallucinations, including hallucinated references, has decreased[1] they have not been entirely eliminated. The persistence of poor or even fraudulent academic work has led to student essays and, in some cases, academic papers, being submitted with fabricated bibliographic references. Much of the literature has focussed on the general prevalence of hallucinations[2-4] and the implications these have for academic integrity.[5,6]

These hallucinated references come in three forms: books, book chapters and journal articles. A common denominator of genAI hallucinated journal references that they are constructed of four genuine or semi-genuine parts: genuine authors (in genuine or semi-genuine combinations), semi-genuine titles, genuine journals and semi-genuine year and volume numbering as well as hallucinated pagination. In some cases, the year of publication corresponds with the correct volume numbering. The pagination is commonly spurious, but may correspond with a correct pagination of existing papers (or different titles and/or subject matter) GenAI hallucinated book references are made of four genuine or semi-genuine parts: genuine authors, years, semi-genuine titles, and genuine publishers. On occasion a place of publication was added. Hallucinated references to book chapters include, in addition, genuine editors (in genuine or semi-genuine combinations) and semi-genuine titles as well as hallucinated pagination.[1]

Given the variables, the number of possible hallucinated journal creations is theoretically infinite. In practical terms it will be constrained, however, by the sources of tokens that are available to the genAI model, which are either based on training or on data ingested from the Internet. These constraints are the number of journals in the relevant scientific discipline, the number of author names that are (well-)published in the field, and the number of keywords or relevant words in the titles of the journal articles.

On the other hand, because these hallucinations occur as part of a text predictive model, it is theoretically also possible, that a genAi model will regenerate an identical hallucinated reference in a different run/chat. At present, no empirical data exist to formally ascertain the existence and potential prevalence of the latter.

Likewise, there is appears to be no clear explanation in the literature how genAI models arrive at hallucinated citations beyond the basic concepts of genAI functionality. This paper will approach the topic in three ways.

It will first examine a set of hallucinated references with the same journal, book and book chapter title to i) illuminate any patterns in these references, ii) whether some of these are duplicated and iii) whether these have been onwards cited.

Work by Ben Williamson has highlighted the existence of a hallucinated journal article attributed to him and his co-author Nelli Piattoeva: "Education Governance and Datafication". Williamson was able to compile 58 references to the non-existing paper.[7] Some of these had additional subtitles added. Given the number of unique citations with the same title, this hallucinated reference was an ideal candidate to examine the generation of hallucinated citations over time.

The second approach will be i) to query ChatGPT itself about the processes it uses to arrive at citations when unable to search the web, and ii) to analyse the responses.

The third approach will be to task ChatGPT with writing a 1000-word referenced essay on a topic related to Datafication and Education Governance and to examine the nature of the sources it cites.

## Methodology

### *Citations encountered in published papers*

The data utilized in part one of this study are derived from three sources:

1. a systematic search in GoogleScholar for the title: "Education Governance and Datafication" (Figure 1). Every source citation was downloaded (using Google Scholar's citation tool)

2. the resulting list of papers was cross-checked against the papers already identified in Williamson's paper[7] with duplicates removed. All residual papers were accessed individually and their citations completed (Williamson does not list the authors).

3. a systematic search using Google for the phrase: "Education Governance and Datafication". All found papers were cross-checked against the papers already identified, with all new items added.

Wherever possible (a small number was behind non-institutional paywalls), the cited source papers were accessed and the full 'Education Governance and Datafication' citation extracted. Where the source paper had been cited by other papers (according to GoogleScholar), each citing paper was also accessed and assessed whether the 'Education Governance and Datafication' paper had been re-cited. The identified source papers and the associated 'Education Governance and Datafication' citations have been reproduced in the appendix.

Where GoogleScholar did not provide active links to the cited source paper (or papers citing the source paper), the paper was sought by other means (e.g. Google Search, direct access to journal, Research Gate). The searches were carried out on 22 March (nº 1 and 2) and 23 March 2026 (nº 3).

### *Querying ChatGPT as to its functionality*

ChatGPT 5-mini was chosen as the genAI model to be tested because ChatGPT is the most readily accessible and widely used model with a current user base of 900 million weekly users.[8] It can also be used to simulate a general user as it provides free access with a limited number of tokens.

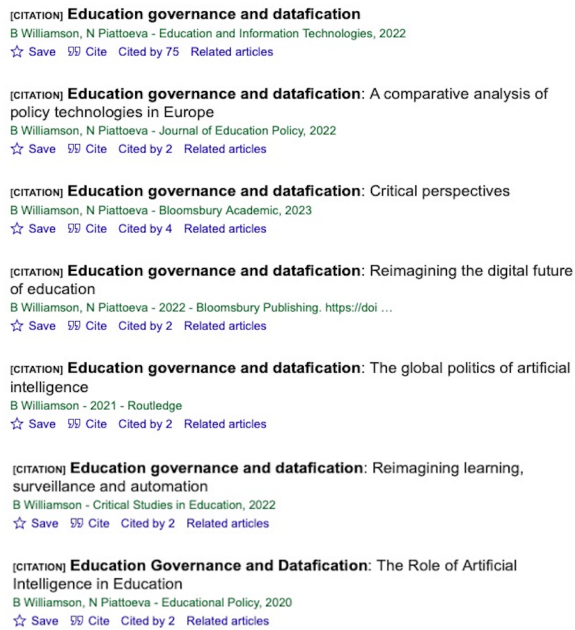

*Figure 1. Screenshots of a Google Scholar search showing citations of Education Governance and Datafication' (31 March 2026)*

Three 'conversations' were carried out with ChatGPT 5-mini via a series of prompts. One conversation focussed on the processes ChatGPT uses to generate citations, esp. citations in an environment where it cannot search the web (Conversation A). The second focussed on generating 250 hallucinated citations and examining the probability of duplications in these (Conversation B). The third examined patterns in the 250 hallucinated citations (Conversation C).

Following the protocol for the reporting of conversations with ChatGPT[9], a supportive document archived at the author's institution provides the full text of the three conversations with ChatGPT that were used and analysed in this paper.[10] In the following text, the individual conversations and prompts are referenced in the format Conversation nº|prompt nº (e.g. A2 for Conversation A, prompt 2).

### References in a ChatGPT-written essay

ChatGPT 5-min was given the following prompt: *"Write a 1000-word essay on the topic of the role of datafication in education in relation to the governance of schools. Cite appropriate references. Use APA6 format."*

The response was copied into a present word file. After saving, the chat was deleted to ensure a clean and unbiased generation of a new essay. In total ten replicated were generated. The essays were created on 31 March 2026 between 08:57 and 09:16 AEDT. Following the protocol for the reporting of conversations with ChatGPT[9], the full text of the essays has been documented in a document archived at the author's institution.[11]

## Part One:
## Analysis of hallucinated citations in the published literature

The bibliographic search resulted in 147 source papers. Of these, 137 papers could be accessed and the hallucinated citations extracted for analysis.

### Date of publication of source papers

The papers contained the hallucinated references under discussion were all 'published' in the past six years, with the bulk of papers published in 2025 (Figure 2). This reflects the perception that recent publications must *per se* be more relevant to the field. After all, citation indices relying on the foregoing two or five years for the evaluation of journals as well as authors.[12-14] Google-Scholar, for example, has the predefined options of restricting literature searches to the current year, the current and previous year, or the past five years.

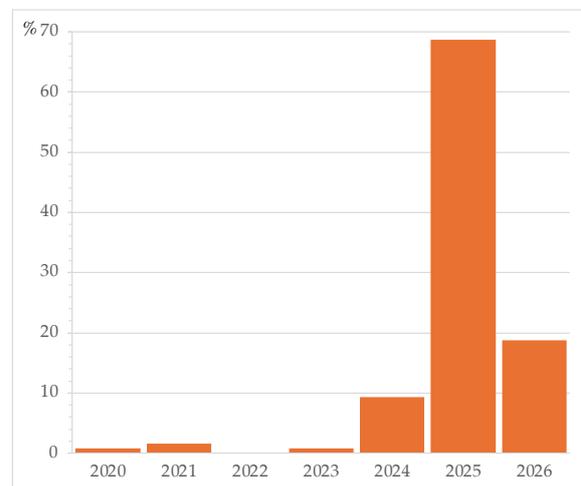

*Figure 2. Frequency of year of publication of source papers (%).*

While ChatGPT will reproduce genuine references to seminal literature [1], its hallucinated references tend to be 'recent.' The majority of hallucinated publication dates were 3 and 4 years prior to the year of the source publication (Figure 3). None of them were older than seven years at the time of publication

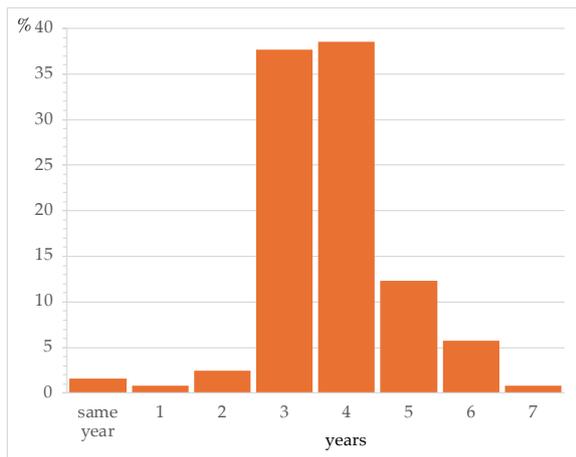

*Figure 3. Age of hallucinated references relative to the year of source publication (n=122).*

### Falsified publication dates

ChatGPT was released to the public in November 2022.[15] Not only did its general usage increase dramatically to a current user base of 900 million weekly users,[8] but from March/April 2023 onwards AI-generated text began to appear in academic writing.[16] Three source papers identified for this study carry dates that predate the public release of ChatGPT and therefore raise suspicion: nº 23, 93 and 112.

Paper nº 93 was purportedly published in the *International Journal for Research in Education* in November 2020. An examination of the journal's website active at the time of writing[1] showed that the link to previously published papers is not working. A perusal of the state of the journal's website on 28 July 2021 (via Internet Way Back Machine)[2] allowed access to previous issues, but only issue vol. 9(4) was listed for 2020.[3] Systematic, year-by-year searches for the paper via GoogleScholar found that it was not indexed until 2025, even though at paper published in vol. 9(4) was indexed in 2022 and several papers published in 2023 were indexed in the same year. Based on the available evidence it appears that paper nº 93 was written in 2025 but back-dated to 2020 by the journal editor(s).

Paper nº 23 differs as systematic, year-by-year searches for the paper via GoogleScholar suggest that was indeed published in 2021. Yet, a year-by-year search of Google search reveals that the title is not indexed until 18 November 2023.[4] Both URLs found in a Google search[5] returned a 404 error and neither was indexed on the Wayback Machine. An Indonesian archive of digital publications provides an entry (including abstract) to the paper, but the downloads links point to the same, now invalid, URL.[6] The journal's home page as indexed by the Wayback Machine does not provide a link to the relevant journal issue in September 2022[7] but was indexed in November 2023,[8] although the relevant page for vol. 12(2) was not archived (first indexed in December 2025).[9] From the fragmented evidence kit appears that the source document would have been written and published in late 2025.

Paper nº 112 on the other hand was reputedly published on 25 December 2021.[10] Inspection of the metadata shows that the paper was supposedly created on that date.[11] Yet the pages for the relevant journal issue, as indexed by the Wayback Machine do not list the paper until 8 July 2025.[12] It would appear that Google draws its on indexing data from on the metadata of the web page.

Given the observations, the single publication dated to 2023 (nº 80) was also assessed for its veracity. It would appear that this date is in fact genuine as the file was published on the Zenodo platform on 28 June 2023.[13] This suggests that this is, in all probability, the following the first documented appearance of the hallucinated reference under discussion. That reference was represented as follows:

> Williamson, B., & Piattoeva, N. (2020). Education Governance and Datafication: The Global Emergence of Data-Driven Decision-Making in Education. *Global Studies of Childhood*, 10(2), 108-120

Digital forensic analysis has shown that none of the asserted dates of the three source papers seeming that predate the public release of ChatGPT3 are true. One wonders as to the rationale for the back-dating such papers. It can be surmised that it relates to issues of perceived primacy of ideas.

### Composition of hallucinated references

Books and book chapters make up a quarter (25.3%) of the hallucinated references (Table 2). Among the 'cited' journals, *Learning, Media and Technology* dominates with 19.9% of all hallucinated references, followed by *Education and Information Technologies* (9.6%) and the *European Educational Research Journal* (8.9%). The remaining 16 journals as well as an encyclopedia ('*Oxford Research Encyclopedia of Education*') are represented with less that 5% each. Four of the 'cited' journals (and encyclopedia) are fictitious, they others do exist (verified via SCImago [17]).

Overall, 89.9% of year and volume combinations (n=89) for cited journals are correct. In the case of the ten instances where this is not the case, all two of the six 'cited' journals was non-existent (verified via SCImago [17]). Of ten 'false' citations, six were published in 2024, three in 2025 and one in 2026.

Table 1. Correlation between the year of publication of a source and the cited hallucination

| hallucinated year | year of publication of source | | | | | | n |
|---|---|---|---|---|---|---|---|
| | 2020 | 2021 | 2023 | 2024 | 2025 | 2026 | |
| 2019 | | | | 1 | 3 | 1 | 5 |
| 2020 | 1 | 1 | 1 | 7 | 18 | 4 | 32 |
| 2021 | | 1 | | 3 | 29 | 1 | 34 |
| 2022 | | | | | 47 | 17 | 64 |
| 2023 | | | | | 3 | | 3 |
| no year *) | | | | 1 | 6 | 1 | 8 |
| n | 1 | 2 | 1 | 12 | 106 | 24 | 146 |

*) hallucinated citations verified as original work could not be assessed

Table 2. Correlation between the year of publication of a source and the cited hallucination

| Title | year of publication of source | | | | | | n |
|---|---|---|---|---|---|---|---|
| | 2020 | 2021 | 2023 | 2024 | 2025 | 2026 | |
| *British Journal of Sociology of Education* | | | | | 1 | | 1 |
| *Critical Studies in Education* | | | | | 4 | 1 | 5 |
| *Education and Information Technologies* | | | | | 5 | 9 | 14 |
| *Education Policy* | | | | 2 | | | 2 |
| *Educational Philosophy and Theory* | | | | 2 | 2 | 1 | 5 |
| *Educational Policy* | | | | 4 | | | 4 |
| *European Educational Research Journal* | | 1 | | | 12 | | 13 |
| *European Journal of Education* | | | | | 1 | | 1 |
| *Global Studies of Childhood* | | | 1 | | | | 1 |
| *Globalisation, Societies and Education* | | | | | 1 | | 1 |
| *Int. J. of Education Policy and Leadership* | | | | | 1 | | 1 |
| *Int. J of Educational Research* | | | | | 1 | | 1 |
| *Int J Educational Technology in Higher Education* | | | | | 2 | | 2 |
| *Journal of Education Policy* | | | | | 4 | | 4 |
| *Learning, Media and Technology* | | | 1 | | 25 | 3 | 29 |
| *Media and Technology* | | | | | | 2 | 2 |
| *Oxford Research Encyclopedia of Education* | | | | | 4 | 2 | 6 |
| *Oxford Review of Education* | | 1 | | | 2 | | 3 |
| *Policy Futures in Education* | | | | | 4 | 1 | 5 |
| *Research in Education* | | | | | 1 | | 1 |
| Book | | | | 2 | 21 | 3 | 26 |
| Book Chapter | | | | | 10 | 1 | 11 |
| no data *) | | | | 1 | 6 | 1 | 8 |
| n | 1 | 2 | 1 | 12 | 106 | 24 | 146 |

*) hallucinated citations verified as original work could not be assessed

While the year and volume combinations for the cited journals tend to be correct, the cited paginations do not exist in the cited or any other volumes of the journals. Spot checks via Google Scholar showed that the cited page ranges do exist among some journals, albeit of very different disciplines.

In total eleven hallucinated citations are comprised of book chapters, all which derived from source paper written in 2025. Among the 'publishers,' Springer looms large (5 instances), followed by MIT Press (2). Bloomsbury Academic, Oxford University Press, Palgrave Macmillan and Routledge are represented with one book each. Eight these eleven citations are verifiable in terms of editors, volume name and publisher. Three are fully hallucinated (nº 114, 136, 141). Among the editors Micheal A. Peters is most commonly mentioned with four (verifiable) entries.

The authorship of the work is overwhelmingly attributed to Ben Williamson and Nelli Piattoeva (93.7%), with seven instances attributing it just to Williamson (nº 26, 35, 47, 55, 98, 106, 121,) and one attributing the title to Ben Williamson and Rebecca Eynon (nº 99).

*Generation of identical references*

In ten instances the same or very similar citations occur in different source papers:

1.  Twelve sources (nº 7, 8, 44, 51, 52, 57, 59, 60, 67, 83, 92, 113) dating to 2025 (4 instances) and 2026 (8 instances) carry the title without subtitles as well as the same journal name, volume number and pagination (*Education and Information Technologies* 27, 3515–3531). In addition, reference sources nº 38 (missing the year) and nº 66 (missing the volume number) are fundamentally the same.

2.  Five sources (nº 22, 46, 72, 95, 96) dating to 2025 (4 instances) and 2026 (1 instance) carry the title as well as the same journal name, year, volume number and pagination (*Learning, Media and Technology* 47(1), 1–15) but differ in their subtitles: 'AI learning analytics and the rise of algorithmic regulation' (nº 72), 'Critical perspectives' (nº 22 'Shifting roles and ethical challenges of AI-based assessment' (nº 95), 'The new politics of knowledge' (nº 96) and 'The Role of AI in Reshaping Learning '(nº 46). In addition, in two source papers (A1 & A2) the data search encountered a paper with a very similar title and identical journal, volume number and pagination ('Education governance and data infrastructures in the age of ai.' *Learning, Media and Technology* 47(1), 1–15).

3.  Reference sources nº 84 and 102 (both published in 2025) carry the title without subtitles as well as the same journal name, volume number and pagination (*Policy Futures in Education* 19(1), 5–20).

4.  Reference sources nº 76 and 77 (both published in 2024) carry the title, as well as the same journal name, year, volume number and pagination ('Education governance and datafication: A comparative analysis of policy technologies in Europe.' *Education* Policy 34(3), 479–499).

5.  Reference sources nº 28 and 79 (both published in 2025) carry the title and the same subtitle as well as the same journal name, volume number and pagination ('Education governance and datafication: A comparative analysis of policy technologies in Europe.' *Journal of Education Policy* 37(2), 194–213).

6.  Reference sources nº 15, 128 and 135 (all published in 2025) carry the title as well as the same journal name, year, volume number and pagination (*Critical Studies in Education* 63(1), 1–16). In addition, references nº 128 and 135 carry the subtitle 'Reimagining education policy through machine intelligence.' A third reference (nº 37) varies in subtitle and pagination (ending in 17).

7.  Reference sources nº 19, 32 and 99 (all published in 2025) have the same journal name, year, volume number and pagination (*Learning, Media and Technology* 45(1), 1–14) but one differs in the authors (Williamson & Enyon in nº 103) and subtitles (none for nº 103, 'A critical analysis' for nº 19 and 'Critical issues and future directions' for nº 34).

8.  Six reference sources (nº 26, 55, 98, 106, 119, 145) (five published in 2025, one in 2026) have the same book details with Williamson as the single author, the same title and the same publisher. Reference nº 98 carries the additional subtitle 'The global politics of artificial intelligence.'

9. Five reference sources (nº 30, 53, 71, 82, 85) carry a book reference with the same title and subtitle ('Education governance and datafication: the power of data in education'), the same year and the same publisher. In two instances (nº 30 and 82) a place of publication is added.

10. Three reference sources (nº 53, 71, 85) (one published in 2024, two in 2025) have the same books details with Williamson and Piattoeva as authors, the same title, the same subtitle and the same publisher.

The overall incidence of duplication is 29.5% (n=146) with duplication of journal references at 28.7% (n=101) and duplication of books and book chapters 37.8% (n=37). The differences between the two proportions are not statistically significant.

### Origin of titles and 'cited' journals

The frequency with which a journal title is included in the hallucinated citations shows no correlation with the journal's H Index (Figure 4), SCImago SJR Index (Figure 5) or the total number of citations articles published in the journal received between 2021 and 2024 (Figure 6).

Searches of relevant literature databases Ebsco,[18] ERIC,[19] Ingenta Connect,[20] Primo,[21] Scopus[22] and Web of Science[23] showed that there is no genuine paper that carries in its title all three key works of the hallucinated title: keywords 'Education,' 'Governance,' and 'Datafication.'

A search of the Web of Science database[23] returned 171 items with that contain three keywords 'Education,' 'Governance' , 'Datafication' either as a keywords or in the full text. The papers are associated with 100 author names. When drilling down, only seven papers carry all three terms as author defined keywords and 36 papers carry all three terms in the abstract. None of these papers were authored either by Williamson or Piattoeva.

The source prompting the hallucination may been this book chapter in an edited volume called 'The Datafication of Education":

> Williamson, Ben, and Piattoeva, Nelli (2020). Objectivity as standardization in data-scientific education policy, technology and governance. In J. Jarke & A. Breiter (Eds.), *The datafication of education* (pp. 81-93). Abingdon: Routledge.[24]

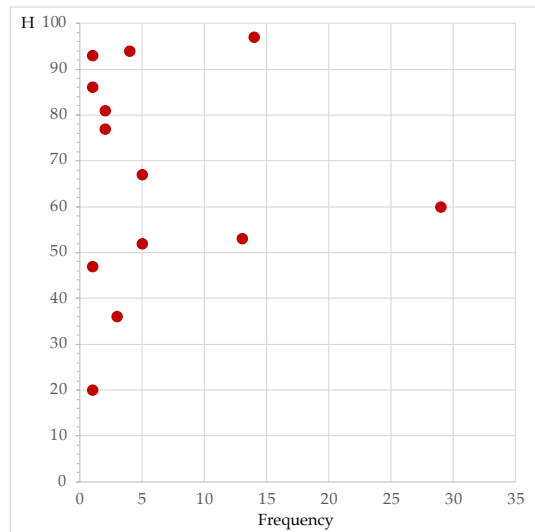

*Figure 4. Relationship between a journal's H-Index and the frequency among hallucinated citations.*

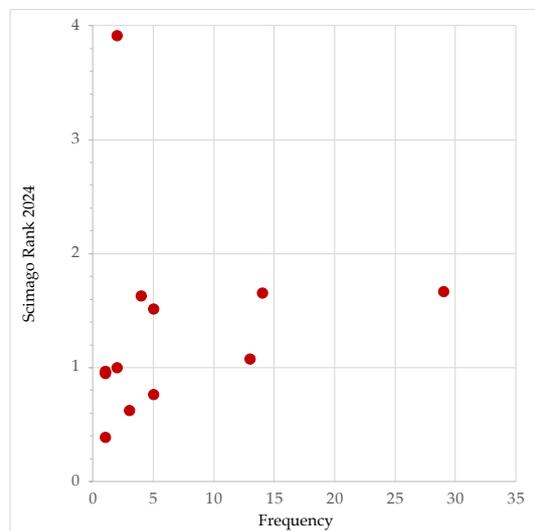

*Figure 5. Relationship between the journal's SCImago Rank 2024 and the frequency among hallucinated citations.*

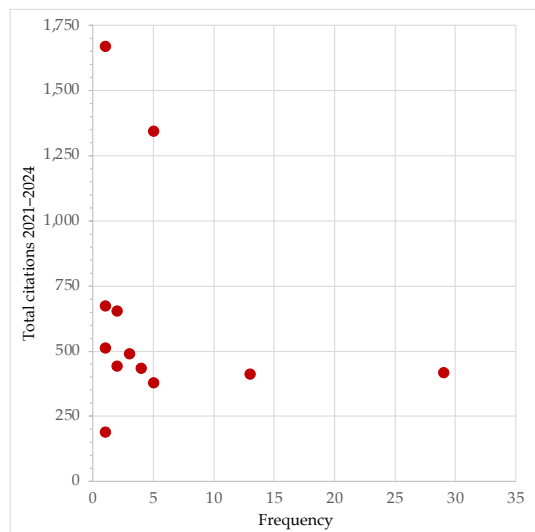

*Figure 6. Relationship between the journal's total citations 2021–2024 and the frequency among hallucinated citations.*

This chapter, which attracted 198 citations (according to GoogleScholar) was 2020 republication of this 2019 paper:

> Williamson, Ben, and Piattoeva, Nelli (2019). Objectivity as standardization in data-scientific education policy, technology and governance. *Learning, Media and Technology* 44(1), 64-76.[25]

Intriguingly, *Learning, Media and Technology* is also the most frequently mentioned journal among the hallucinated references, and it is the one where the paper originally was published

While Piattoeva had published 21 papers with the word datafication and Williamson had published 103 such papers,[23] the above-cited paper is the only one that Williamson and Piattoeva published together.[23]

### Onward citation

Eighteen of the 128 source papers had been cited by other authors, ranging from one to fifteen (nº 88) citations. All such citations were individually assessed to ascertain whether the paper 'Education Governance and Datafication' was also cited in these. No second-generation citations were detected.

Searches of the literature databases Ebsco,[18] ERIC,[19] Ingenta Connect,[20] Primo,[21] Scopus[22] and Web of Science[23] also showed that a 'paper' with the title 'Education Governance and Datafication' was not listed.

While the hallucinated citation 'Education Governance and Datafication' has been included as [Citation] in GoogleScholar (which served as the source for part of the data extraction for this paper), this has not (yet) been added to either Ben Williamson's nor Nelli Piattoeva's GoogleScholar profile. Thus, these hallucinated citations do not (as yet) inflate the authors' citation indices.

# Part Two:
# ChatGPT and its hallucinated citations

Given that none of the authors of the source papers attested to the use of genAi (let alone specific models) in the creation of their manuscripts, we can only surmise that ChatGPT, the most freely accessible widely available tool, would have been the AI model that generated the text and the associated hallucinated references.

### Background

Large Language Models (LLMs) are advanced neural network architectures designed for natural language understanding and generation. Typically based on the transformer architecture, they leverage self-attention mechanisms to capture long-range dependencies and contextual relationships across tokens in text sequences. During pretraining, LLMs are optimized via next-token prediction over massive corpora, enabling them to internalize syntactic structures, semantic relationships, factual knowledge, and domain-specific patterns. Fine-tuning or instruction-tuning further adapts these models to specific tasks, improving coherence and alignment with user objectives. At inference, LLMs generate probabilistic token sequences conditioned on input prompts, employing various techniques to balance creativity and relevance. While LLMs exhibit emergent capabilities in reasoning and context-aware generation, they operate through statistical pattern recognition rather than genuine understanding or cognition, making their outputs highly sensitive to training data distribution and prompt formulation.

### The conversation

The following section is based on a 'conversation' with ChatGPT 5-mini (henceforth ChatGPT), the full text of which can be accessed at the author's digital repository.[10] That 'conversation' followed a structure approach by asking what the genAI model know about the author Ben Williamson with the specific instruction of not searching the web. This was followed by a prompt asking for a concise list of works, specifically requesting that no web search be conducted (A2). The aim of this approach was to discover what level of information had been ingested by the model during its training phase. Normally, direct and specific queries about the nature of the ingested training materials are rejected by the model (Box 1). When embedded in the prompt sequence, however, ChatGPT generated six references, five of were hallucinated references. One, however, was genuine and was cited with the correct year ('Big Data in Education: The Digital Future of Learning, Policy and Practice' 2027).[26] When the prompt was rerun later on (A4), ChatGPT again generated reference to the genuine book as well as five of hallucinated references.

*Box 1: Response by ChatGPT 5-mini when queried about a specific work (26 March 2026, 03:09 GMT).*

| DHRS | Have you ingested the following title during your training phase? "Big Data in Education: The Digital Future of Learning, Policy and Practice" |
|---|---|
| ChatGPT 5-mini | I don't have access to my training data, so I can't confirm whether a specific book like "Big Data in Education: The Digital Future of Learning, Policy and Practice" was included in it. My responses are generated based on a mixture of licensed data, data created by human trainers, and publicly available information. |
| | I can provide summaries, insights, or analyses of its content if you want, based on general knowledge of big data in education and trends in learning, policy, and practice. |
| | Do you want a brief overview of what this book likely covers? |

When queried, ChatGPT responded that its answer was based on pattern recognition from texts, synthesis of public knowledge (bases on references to Williamson's work ChatGPT had seen), and inferences from context (A5). Critical is the assertion that "everything I share without searching comes from [my] static knowledge base" (A5). When asked regarding the work 'Education Governance and Datafication' which it had listed in a response (A4), ChatGPT referred to this as a description of a cluster of Williamson's works (A6). It offered to provide a list of actual titles, two of which were genuine[26,27] and three were hallucinated.[14]

Subsequent prompting ascertained that (most?) citations are reconstructed based on patterns in memory. The conversation then moved to the mechanics of reference reconstruction (A9–A12) and whether ChatGPT would view source reference n° 9 as genuine reference without being able to verify this via web searches (A13-14) (which it asserted). It asserted it as genuine, but when allowed to search the web identified it as non-existent (A15). The conversation then moved on to whether that hallucinated citation could have been generated by ChatGPT (A16), how the process of hallucinated citations worked (A17–A19) and the probability of duplicate hallucinated citations being generated (A20–A28).

*Process of reference (re-)construction*

According to ChatGPT's response (A9), a reference it provides is not a random fabrication but is reconstructed based on patterns derived from training data (e.g. academic papers, book chapters, abstracts, citation lists, bibliographies and literature reviews); memories of widely cited and discussed works in conceptual fields; and logical inferences from context (e.g. co-authors, journals in the research field, approximate years in relation the author flourishing).

Using the title 'Education Governance and Datafication' as an example (in the context of the conversation that was grounded in a query on Williamson's work), the (re-)construction process looked at i) the recognition of themes which identified both 'education governance' and 'datafication' as concepts that appear repeatedly across his papers and discussions of his work; ii) pattern generalisation by combining the two key concepts according to typical academic phrasing (titles combine the research object + method/approach + policy context)

*Generating hallucinated citations*

When a user asks for academic references or citations, ChatGPT will generate plausible-sounding references by matching the topic with authors known to be working in this field (favouring well cited authors by virtue of representation in the training database); with journals well cited in the field and attaching plausible years volume/issue numbers, and page ranges. The page ranges are not memorised exact values but plausible numbers derived from patterns of typical volumes, issues, and years for that journal. Among these 'core' fields (authors, main title, journal) are highly stable while peripheral fields (subtitles, volume/issue, pages) are probabilistic (A28).

In consequence, the model produces a reference that looks real, is stylistically correct, and fits the topic—but does not exist. As the "the model is designed to produce fluent, coherent text, it sometimes results in plausible-but-fake references" derived from a programmed aim of "provide a helpful and credible-looking reference" (A17). The probability of generating a duplicate reference (bar the pagination) is $p$ =0.055 (i.e. 5.5%) (A20), while the inclusion of page numbers drops the probability to $p$= 0.000675 (i.e. 0.05%).[15]

Conversation B6 outlines a hypothetical scenario how a hallucinated reference can enter the formal citation universe. While this shall not be repeated here in detail the process is facilitated by cognitive biases among time-poor or not fastidious users, namely: authority bias (recognizable and reputable author sand journals); confirmation bias (the reference fits pattern of other work of the authors); plausibility (perfect formatting masks non-existence) and repetition bias (appears to be genuine as the reference was seen multiple times).

### *Experimentally generated hallucinated citations*

In an experiment, ChatGPT generated five sets of 50 citations along the lines of Williamson's work (B1–B5). Among these, the titles 42 references (16.%) were duplicated, with one reference repeated seven times, seven references four times and 13 references three times. The remainder of the duplicates occurred twice.

All 176 unique generated titles were assessed via whether they existed as citations 'in the wild.' Three of these could be found. The title 'Education governance and data infrastructures' exits in two citations,[16] while the titles 'Datafication and education governance'[17] and 'Governing education through datafication'[18] have been cited once. In addition, once source lists a version of the title 'Datafication and education governance', also in a hallucinated citation, albeit attributed to authors other than Williamson.[19]

## Part Three:
## References in a ChatGPT-written essay

Ten replicate essays were generated with the following prompt given to ChatGPT 5-min: "*Write a 1000 word essay on the topic of the role of datafication in education in relation to the governance of schools. Cite appropriate references. Use APA6 format.*"

The ten essays 'cited' a total of 98 references, 57 of which were different. Of these, two thirds (66.7%) are genuine and 17.5% were genuine with errors. As the errors among the latter group may have been derived from faulty citations that ChatGPT could accessed on the web, they had been classified as 'genuine with errors.' Fully hallucinated were 3.5% of the citations, while the remaining 12.3% had sufficient numbers of serious errors that they cannot be the result of genuine web searches.[11]

Among the two hallucinated references was one (nº 56)[11] with the exact same characteristics as the hallucinated reference under discussion earlier in this paper:

> Williamson, B., & Piattoeva, N. (2022). Education governance and datafication. *Education and Information Technologies*, 27, 3515-3531.

As the generation of referenced essays has shown, ChatGPT-written essays will continue to carry hallucinated references even though the model has access to the web. The ten essays that had been generated still contained 9.2% hallucinated references. Particularly intriguing is that one of these was an exact match of the hallucinated references that was most commonly observed among the published source papers (see p. 6).

## Implications

The hallucinated academic references created by ChatGPT are not random errors but predictable, pattern-driven artifacts of how genAI models generate text. These references are systematic reconstructions built from real authors, journals, and topical keywords. Because genAI models rely on pattern recognition rather than factual verification, they produce citations that are structurally correct and contextually plausible, yet non-existent. These hallucinations can be repeatedly generated, leading to duplication and consistency across different texts, which can increase their perceived legitimacy.

The ability for genAI models to search the net is crucial in the suppression of hallucinated references. Ideally, hallucinated should be a thing of the past. ChatGPT was web search enabled for the users of the subscription service (ChatGPT Plus and ChatGPT Team) from 31 October 2024 onwards. Web search was rolled out for all users from 16 December 2024.[28] As the generation of referenced essays has shown, however, ChatGPT-written essays will continue to carry hallucinated references even though the model has access to the web.

At present, no second-generation citation of the examined hallucinated reference has been found. Given that citation laundering (i.e. citing a source because others cited it) is not uncommon, it may well be only a matter of time that such second-generation citations will occur.

# Endnotes

1. http://www.raijmr.com/ijre.
2. Link: https://web.archive.org/web/20210728215716/http://www.raijmr.com/ijre/vol-9-issue-4-april-2020.
3. For 2021, only the issue vol.10(5) was listed.
4. It first appears on the author's Google Scholar profile page.
5. https://sttbi.ac.id/journal/index.php/edukasi/en/article/view/538.— https://sttbi.ac.id/journal/index.php/edukasi/en/article/view/538/277.
6. https://garuda.kemdiktisaintek.go.id/documents/detail/5528273.
7. The most recent issue listed on that archived page was vol.11(1), 2020.
https://web.archive.org/web/20220927003400/https://sttbi.ac.id/journal/index.php/edukasi/issue/archive
8. https://web.archive.org/web/20251228163630/https://sttbi.ac.id/journal/index.php/edukasi/article/view/538/277
9. https://web.archive.org/web/20251228163630/https://sttbi.ac.id/journal/index.php/edukasi/article/view/538/277
10. See article overview page: https://rigeo.org/menu-script/index.php/rigeo/article/view/2868
11. "DC.Date.created= DC.Date.created" scheme="ISO8601" content="2021-12-25.—But see "DC.Date.created" scheme="ISO8601" content="2021-12-25 " and "DC.Date.dateSubmitted" scheme="ISO8601" content="2024-12-29
12. Not listed for 9 February 2023, 27 February 2024, and 22 July 2024. Listed for 8 July 2025.—
https://web.archive.org/web/20230209084844/https://rigeo.org/menu-script/index.php/rigeo/issue/view/47 —
https://web.archive.org/web/20240227030458/https://rigeo.org/menu-script/index.php/rigeo/issue/view/47 —
https://web.archive.org/web/20240722083353/https://rigeo.org/menu-script/index.php/rigeo/issue/view/47 —
https://web.archive.org/web/20250708202619/https://rigeo.org/menu-script/index.php/rigeo/issue/view/47.
13. https://zenodo.org/records/16014628.
14. Intriguingly, one of the hallucinated citations produced in A7 was found to cited in thirteen publications attesting to proliferation of ChatGPT generated hallucinated citations:
Williamson, B., & Piattoeva, N. (2019). "Objectivity as standardization in data-scientific educational governance: Grasping the global through the local." Research in Education, 101(1), 69–91
15. Rerun of prompts A17 to A20, with specification to include page numbers (page ranges which are considered especially noisy as genitive AI models tend to produce 'reasonable-looking' but not consistent spans).
16. Cited in: Qureshi, S., & Irshadullah, H. M. (2025). From AI Ethics to AI Justice: A Comprehensive Framework for Equitable Governance in Education and Social Systems. VFAST Transactions on Education and Social Sciences, 13(3), 55-72. https://vfast.org/journals/index.php/VTESS/article/download/2264/1789/10673. Citation: as Williamson, B., and Piattoeva, N. (2022). Education governance and data infrastructures in the age of ai. Learning, Media and Technology, 47 (1), 1–15.— It was cited twice in : Sitorus, J. (2025). Transformasi Evaluasi Pembelajaran Berbasis Kecerdasan Buatan: Peluang dan Tantangan Masa Depan. Immanuel: Jurnal Teologi dan Pendidikan Kristen, 6(1), 229-241. https://stt-su.ac.id/e-journal/index.php/immanuel/article/view/340.—cited as Williamson, B., & Piattoeva, N. (2023a). Education governance and data infrastructures in the era of. Learning, Media and Technology, 1. and as Williamson, B., & Piattoeva, N. (2023b). Education governance and data infrastructures in the era of AI: Ethical challenges and policy responses. European Educational Research Journal, 2.
17. Cited in: Guedes, Felipe (2026) O Impacto da Inteligência Artificial na Educação Contemporânea: Desafios e Perspectivas.
https://www.scribd.com/document/947802919/Someon.—Cited as: Williamson, B., & Piattoeva, N.. Datafication and Education Governance: Critical Perspectives. New York: Routledge, 2022..
18. Cited in: Al-Sharari, Mufadhi Ratyan (2022). Educational Leadership and the Integration of Artificial Intelligence in Higher Education: Strategies, Challenges, and Opportunities, Educational Administration: Theory and Practice, 28(3) 494-504 Doi: 10.53555/kuey.v28i03.10517—Cited as: Gulson, K. N., Sellar, S., & Williamson, B. (2018). Governing education through datafication: Sundering learning and equity. Educational Philosophy and Theory, 50(6–7), 581–591. https://doi.org/10.1080/00131857.2017.1372639.

[19]. Cited in the blog 'When Evidence Isn't Enough' (https://www.edblogcast.com/when-evidence-isnt-enough/). Cited as Lewis, S., & Hartong, S. (2022). From data to decisions? Exploring the datafication of education governance. Educational Policy, 36(1), 3–26. https://doi.org/10.1177/0895904820983037.

# References

# Appendix

This appendix sets out the full citations of the identified source documents as well as the hallucinated citations as cited in the source paper. The author is conscious that there may be structural issues why the authors of the source papers may have chosen to rely on generative AI tools to support their writing. Institutional pressures along the lines of publish or perish may have forced authors to take short cuts. A lack of AI literacy may well have resulted in unwarranted faith being placed in an output that was deemed to be genuine and trustworthy. Academic integrity nonetheless requires that authors verify their sources and where they cannot, either to refrain from citing them or to cite them with the appropriate caveats of 'not seen' or 'cited after.'

| Ref | Reference | Hallucinated Citation |
|---|---|---|
| 1 | Abduqodir, Mahamatnabiyeva Barno (2025). GEN-Z Slangs And The Ai Revolution In Language Teaching Journal of Applied Science and Social Science Volume 15 Issue 10, October 2025, pp762– | Williamson, B., & Piattoeva, N. (2021). Education governance and datafication. Oxford Research Encyclopedia of Education. |
| 2 | Achermann, K. (2025). Artificial Intelligence in Education: A Multidisciplinary Assessment of Learning Outcomes, Pedagogy, Ethics, Accessibility, and Policy. Canadian Journal of Multidisciplinary Research, 1(1), 37-40. | non verifiable, link does not respond |
| 3 | Adalihan, N., Skripnikova, A. I., & Shyngyssova, N. T. (2025). Issues of journalistic ethics in the context of generative artificial intelligence: based on material from Kazakhstan and China. Bulletin of LN Gumilyov Eurasian National University. JOURNALISM Series, 153(4), 137-149 | Williamson B., Piattoeva N. Education governance and datafication: Beyond policy borrowing. Globalisation, Societies and Education. 19(3), 243-260(2021). DOI: https://doi.org/10.1080/14767724.2020.1864755 |
| 4 | Akela, S. R Reimagining Education: The Transformative Impact Of Artificial Intelligence On Teaching And Learning." Reimagining Research: 42. 2025. | Williamson, B., & Piattoeva, N. (2021). Education governance and datafication. Policy Futures in Education, 19(6), 680–696 |
| 5 | Akkaliyeva, A., & Ospanova, K. (2025). Practical Use Of Artificial Intelligence In Teaching English To 5th Grade Students. Вестник Шәкәрім Университета. Серия Филология, 3(3), 183-190. | Williamson B., Piattoeva N. Education governance and datafication // Oxford Research Encyclopedia of Education. – Oxford: Oxford University Press. 2022 |
| 6 | Al Samman, A Towards a Smarter University: A Conceptual Framework for Artificial Intelligence Integration in Higher Education." Revista de Educacion, 411(1.1), Jan 2026. | Williamson, B., & Piattoeva, N. (2022). Education governance and datafication. In M. A. Peters (Ed.), Encyclopedia of Educational Philosophy and Theory (pp. 1–6). Springer. https://doi.org/10.1007/978-981-287-532-7_694-2 |
| 7 | Al Yakin, A., Muthmainnah, M., Asad, M. M., & Almusharraf, N. M. (2026). Rethinking sociology education in human–robot social interaction learning ecosystems: a mixed method study. International Journal of Sociology and Social Policy, 1-15. | Williamson, B. and Piattoeva, N. (2022), "Education governance and datafication", Education and Information Technologies, Vol. 27, pp. 3515-3531 |
| 8 | Aldino, A. A., Maheshi, B., Li, Y., Zhou, Y., Tsai, Y. S., Gašević, D., & Chen, G. (2026). Enhancing learner-centered feedback with AI: teachers' practices and perceptions. Assessment & Evaluation in Higher Education, 1-18. | Williamson, B., and N. Piattoeva. 2022. "Education Governance and Datafication." Education and Information Technologies 27: 3515–3531 |
| 9 | AMIN, SNEHAL S. (2025) Ai And The Future Of English As A Second Language Teaching:. A Study With Special Reference To Gujarat State Indian e-Journal on Teacher Education 13(10, 1–16 | Williamson, B., & Piattoeva, N. (2020). Education governance and datafication: The growing influence of artificial intelligence. Learning, Media and Technology, 45(1), 1–15 |
| 10 | ANTIFASCHISTISCHES SCHUTZ-INSTITUT für reflexive Gesellschaftsethik (2025). Antiautoritäres Studium Generale. Systematische Analyse, Kritik und Alternativen für Bildung, Subjektwerdung und gesellschaftliche Teilhabe. | Williamson, Ben; Piattoeva, Nelli (2022): Education Governance and Datafication. Oxford: Routledge |

| Ref | Reference | Hallucinated Citation |
|---|---|---|
| 11 | Anwar, N. (2024). Artificial intelligence in education: Personalized learning and beyond. Frontiers in Artificial Intelligence Research, 1(3). | Williamson, B., & Piattoeva, N. (2019). Education Governance and Datafication: The Politics of Data in the Digital Age. Cambridge University Press |
| 12 | Anwar, N. (2024). Artificial intelligence in education: Personalized learning and beyond. Frontiers in Artificial Intelligence Research, 1(3). | Williamson, B., & Piattoeva, N. (2020). Education Governance and Datafication: The Role of Artificial Intelligence in Education. Educational Policy, 34(5), 1053-1078 |
| 13 | Ardianti, T., Evitarini, A., & Brescia, R. (2025). Evaluasi Pembelajaran dan Program Bimbingan Konseling. Cerdas Akademika Nusantara. | not verified [eBook, not accessible without payment] |
| 14 | Avwerosuoghene P. Akiti (2025). The Application of Artificial Intelligence in the Field of Education: Challenges. https://doi.org/10.5281/ZENODO.17089458 | Williamson, B., & Piattoeva, N. (2019). Education governance and datafication. European Educational Research Journal, 18(6), 701–717. |
| 15 | Ayoví, F. M., Campos, D. V., Romo, V. V., & Curicho, E. G. (2025). Inteligencia artificial y equidad educativa oportunidades y riesgos en la educación pública. Polo del Conocimiento, 10(9), 3262-3277 | Williamson, B., & Piattoeva, N. (2022). Education governance and datafication. Critical Studies in Education, 63(1), 1–16. https://doi.org/10.1080/17508487.2020.1866052 |
| 16 | Bansal, T., Rawat, A., & Gupta, R. (2025, November) The CAIR Model: A Contextualised AI Integration and Reflection Framework for Teacher Empowerment in Education. In 2025 14th International Conference on System Modeling & Advancement in Research Trends (SMART) (pp. 362-370) IEEE. | Williamson, B., & Piattoeva, N. ( 2021 ). Education governance and datafication. Oxford Review of Education, 47 ( 2 ), 184–201. https://doi.org/10.1080/03054985.2020.1835627 |
| 17 | Baowen, L., & Panicker, C. M. V. (2025). Shifting Paradigms of Educational Leadership in The Context of Artificial Intelligence Integration: A Case Study of Chinese Schools." Sciences of Conservation and Archaeology 37.3 (2025).: 35-41. | Williamson, B., & Piattoeva, N. (2022). Education governance and datafication. European Educational Research Journal, 21(3), 319–334. https://doi.org/10.1177/14749041211062167 |
| 18 | Bhaskaru, O., & Lalitha, K. (2026). Balancing AI-Driven Innovation With Authentic Learning in Higher Education. In Empowering Educational Development and Faculty Growth With AI (pp. 71-102) IGI Global Scientific Publishing | Williamson B. Piattoeva N. (2021). Education governance and datafication.Learning, Media and Technology, 46(1), 1–8. |
| 19 | Bình, B. H. V., & Sơn, S. N. Q. (2025). Chuyển Đổi Số Trong Giáo Dục Đại Học Ngoài Công Lập: Từ Đổi Mới Công Nghệ Đến Đổi Mới Sư Phạm. Tạp chí Khoa học Trường Đại học Trưng Vương, 1(3). | Williamson, B., & Piattoeva, N. (2020). Education governance and datafication: A critical analysis. Learning, Media and Technology, 45(1), 1–12. https://doi.org/10.1080/17439884.2020.1718459 |
| 20 | Boell, M. (2025). La relación con el saber y la inteligencia artificial en la educación superior. Trayectorias Universitarias, 11, 1–11 | Williamson, B., y Piattoeva, N. (2023). Education governance and datafication: Critical perspectives on analytics and AI in schooling. Routledge. |
| 21 | Bonura, A., Di Martino, V., & Inglese, G. (2025). La leadership educativa del dirigente scolastico: un modello di valorizzazione delle risorse umane per la crescita della comunità scolastica. Dirigenti Scuola, (44). | Williamson B., Piattoeva N , Education, governance and datafication», in U. Felt –R. Fouché –C.A. Miller –L. Smith-Doerr (edd.),The handbook of science and technology studies, MIT Press, Cambridge (MA) 2022, pp. 1-22. |
| 22 | Bustamante, R. E. P., Carguaquispe, M. H. S., & López, H. T. V. (2026) Inteligencia Artificial, Políticas Educativas e Innovación en la Transformación de la Educación Superior en Ecuador: Una Revisión Sistemática (2015–2025). Polo del Conocimiento, 11(2), 1357-1369 | Williamson, B., & Piattoeva, N. (2022). Education governance and datafication: Critical perspectives. Learning, Media and Technology, 47(1), 1–15. https://doi.org/10.1080/17439884.2021.1950873 |
| 23 | Chandra, D. C., Susanto, G. H., Pangalinan, T., & Ariel, O. Y. P. (2021). Kepemimpinan Digital: Mengelola Institusi Pendidikan Di Era Postdigital. Edukasi, 12(2), April 2021 | Williamson, B., & Piattoeva, N. (2020). Education governance and datafication. European Educational Research Journal, 19(6), 1–7. |
| 24 | Chayati, S. (2025). Pendekatan Deep Learning Sebagai Alat Bantu Pengajaran Interaktif Di Sekolah Dasar. Jurnal Pendidikan Pemuda Nusantara, 7(2), 32-41. | Williamson, B., & Piattoeva, N. (2020). Education governance and datafication. Learning, Media and Technology, 45(3), 1–15. https://doi.org/10.1080/17439884.2020.1691240 |

| Ref | Reference | Hallucinated Citation |
|---|---|---|
| 25 | Cheah, K. S., Tang, S., Kularajasingam, J., Jambulingam, M., & Beh, W. F. (2025) Evaluating teachers' perspectives on education-based assessments strategies with artificial intelligence. International Journal of Education and Practice, 13(3), 923-942. | Williamson, B., & Piattoeva, N. (2020). Education governance and datafication: The role of artificial intelligence in education. Educational Philosophy and Theory, 52(1), 1-12. |
| 26 | Čirić, Nedim. Kriza I Integracija Pedagogije U Doba Obrazovnih Transformacija. Zbornik radova Islamskog pedagoškog fakulteta u Bihaću 17, no. 17 (2025).: 157-17 | Williamson, B. (2021). Education governance and datafication. Routledge. https://doi.org/10.4324/9781003001348 |
| 27 | Collins, M. A., & Thompson, S. E. (2025). Artificial Intelligence Integration to Improve the Effectiveness of Digital Education." Innovate Inovation Teaching And Learning 1.3 (2025).: 1-17. https://journal.jkrisetmetrolampung.web.id/index.php/innovate/article/view/99 | Williamson, B., & Piattoeva, N. (2021). Educational governance and datafication. Learning, Media and Technology, 46(1), 1–4. https://doi.org/10.1080/17439884.2021.1876089 |
| 28 | Coppi, A., Guerrini, V., & Barone, S. L. (2025). L'intelligenza artificiale in ambito scolastico: percezione e utilizzo da parte dei docenti della scuola secondaria. Riflessioni da un'indagine esplorativa." IUL Research 6.12 (2025).: 181-195. | WILLIAMSON,B.,& PIATTOEVA,N.(2022). Education governance and datafication: A comparative analysis of policy technologies in Europe. Journal of Education Policy, 37(2), 194–213. |
| 29 | Costramos, N. (2025). Integrating Artificial Intelligence Tools in Classroom Language Assessment: Uses, Perceptions, and Pedagogical Implications." BRU ELT JOURNAL 3.3 (2025).: 271-283. | Williamson, B., & Piattoeva, N. (2022). Education governance and datafication: The role of AI in assessment policy. Learning, Media and Technology, 47(2), 213-226. https://doi.org/10.1080/17439884.2021.1942957 |
| 30 | da Silva, I. N., & de Souza, J. C. (2025). Inteligência Artificial e Psicologia Cognitiva: contribuições no desenvolvimento de tecnologias educacionais adaptativas na educação básica. Revista Sítio Novo, 9, e1781-e1781. | WILLIAMSON, Ben; PIATTOEVA, Nelli. Education governance and datafication: the power of data in education. London: Routledge, 2021 |
| 31 | Daneva, M. (2025, September) Application of Artificial Intelligence in E-Learning as a Tool for Inclusive Education. In 2025 XXXIV International Scientific Conference Electronics (ET) (pp. 1-6) IEEE. | Williamson B., Piattoeva N, "Education governance and datafication," Journal of Education Policy, vol. 37, no. 1, pp. 1–19, 2022 |
| 32 | Debbarma, Newton & Chowdhury, Arnab (2025). An Overview on the Uses of Artificial Intelligence (AI) in the School Education System. International Journal for Multidisciplinary Research 7(6), 1–6 | Williamson, B., & Piattoeva, N. (2020). Education governance and datafication: Critical issues and future directions. Learning, Media and Technology, 45(1), 1–14. |
| 33 | Delavarpour, P., & Safarnejad, E. (2024). Strategic Approaches to Student Involvement in Formative Assessment: Insights From Qualitative Interviews. Assessment and Practice in Educational Sciences, 2(1), 1-9 | non verifiable, link does not respond The journal is published in Tehran. At the time of writing (end of March 2026), the internet services in Iran were disconnect due to the war. |
| 34 | Derizhan, Itska (2025). Ethical Issues In The Use Of Artificial Intelligence (Ai) In Education. Жаһандық өзгерістер жағдайындағы ғылым мен білімнің өзекті мәселелері атты XII Халықаралық ғылыми-практикалық конференциясының. материалдар жинағы 12-13 қараша 2025 жыл 12–14 | Williamson B., Piattoeva N. Education Governance and Datafication. — Cham: Springer, 2020. — 250 p. |
| 35 | Di Domenico, M., Romano, F., Amendola, G., & Di Tore, P. A. (2025). Education, artificial intelligence and blockchain: a systematic review on the convergence of emerging technologies and learning processes." Journal of Inclusive Methodology and Technology in Learning and Teaching 5.3 (2025). | Williamson, B. (2021). Education governance and datafication: The construction of policy through data. Learning, Media and Technology, 46(4), 435–448. |
| 36 | Didi, N. (2025). Educational Integrity: Ethical and Regulatory Perspectives on Generative AI in Moroccan Higher Education." Confluence Journal of Multidisciplinary Studies (2025).: 208-221. | Williamson, B., & Piattoeva, N. (2020). Education governance and datafication: The role of the state in the digitization of education. International Journal of Educational Technology in Higher Education, 17(1), 1-13 |

| Ref | Reference | Hallucinated Citation |
|---|---|---|
| 37 | Dikmen, S., & Bahçeci, F. (2025). Measuring Students' Instantaneous Attention Levels in Classroom Environment: Development and Evaluation of Artificial Intelligence Software. e-Kafkas Journal of Educational Research, 12(3), 868-882. | Williamson, B., & Piattoeva, N. (2022). Education governance and datafication: The state, digital technologies and education policy. Critical Studies in Education, 63(1), 1–17 |
| 38 | Dimari, A., Tyagi, N., Sohibova, Z., Mirzaeva, N., Sripada, N. K., & Sharma, M. (2025, March) Leveraging Machine Learning to Personalize Learning and Reduce Close Book Examination Stress. In 2025 International Conference on Frontier Technologies and Solutions (ICFTS) (pp. 1-11) IEEE. | Williamson, B., & Piattoeva, N. Education governance and datafication. Education and Information Technologies, 27, 3515–3531 |
| 39 | Dinara, Mukhtarova (2025). The Impact Of Generative Artificial Intelligence On Literature Instruction:Opportunities And Constraints. International Scientific Conference "Artificial Intelligence And Humanitarian Sciences: Modern Situations And Perspectives" 21 Noyabr 2025 Materials 250–251 | Williamson, B., & Piattoeva, N. (2022). Education Governance and Datafication: Critical Perspectives on the Digital Transformation of Education. Learning, Media and Technology, 47(3), 299–312 |
| 40 | Dolfina, F. M., Lastriani, P. G. E., Putri, J. A. L., Priatna, I. K. S. A., Sitepu, F. B., & Sugiantari, K. H. (2025). Convenient or concerning: Navigating the benefits and risks of artificial intelligence in education." Socionomics 1.1 (2025): 23-37. | Williamson, B., & Piattoeva, N. (2022). Education governance and data infrastructures: AI, algorithms and datafication. European Educational Research Journal, 21(3), 351–369. |
| 41 | Dolfina, F. M., Lastriani, P. G. E., Putri, J. A. L., Priatna, I. K. S. A., Sitepu, F. B., & Sugiantari, K. H. (2025). Convenient or concerning: Navigating the benefits and risks of artificial intelligence in education." Socionomics 1.1 (2025): 23-37. | Williamson, B., & Piattoeva, N. (2022), Education governance and datafication. In A Wilkins & A Olmedo (Eds.) Education governance and social theory (pp 92–110) Bloomsbury Academic |
| 42 | Dumbuya, Emmanuel (2025). Artificial Intelligence and Education: Preserving Human Agency in the World of Automation https://files.eric.ed.gov/fulltext/ED664881.pdf | Williamson, B., & Piattoeva, N. (2020). Education governance and datafication: The role of the state in the digitization of education. International Journal of Educational Technology in Higher Education, 17(1), 1-13. |
| 43 | Dutta, M. S. (2024). The Ethics of AI in Teacher Education: Challenges and Opportunities." Pen & Prosperity, 1(2), Dec 2024 | Williamson, B., Piattoeva, N. (2020). Education governance and datafication: The role of data in educational policy-making. *Educational Policy*, 34(4), 617-644. https://doi.org/10.1177/0895904818807763 |
| 44 | Elsaiary, R. (2026). Preparing Educational Leaders for AI Literacy Through Ethical, Purpose-Driven, and Culturally Responsive Leadership. In Transforming Policy and Practice Through AI-Driven Educational Leadership (pp. 105-132) IGI Global Scientific Publishing | Williamson B. Piattoeva N. (2022). |
| 45 | Fadil, H. O. C. A., & Nuredin, A. (2025, January) Algorithmic bias in AI-enhanced education: Cultural dimensions and pedagogical impact. In International Symposium On Law, Justice, And Emerging Global Challenges(P. 163). | Williamson, B., & Piattoeva, N. (2022). Education governance and datafication. European Educational Research Journal, 21(2), 143-160. |
| 46 | Ferdosipour, A. (2025). Applying Cognitive Psychology Theories In Education In The Era Of Artificial Intelligence. Proceedings 10th International Scientific Conference The Current Issues in Theoretical and Applied Psychology Yerevan, Armenia October 10-12. 2025. Yerevan State University, Yerevan 272-274 | Williamson, B., & Piattoeva, N. (2022). Education Governance and Datafication: The Role of AI in Reshaping Learning. Learning, Media and Technology, 47(1), 1–15. |
| 47 | Gandawa, G. (2026). The Autonomous University: AI-Driven Governance, Admissions, and Academic Planning Author." Journal of Education and Learning Sciences (2026). | Williamson, B. (2022). Education governance and datafication: The rise of digital platforms and algorithmic regulation. Polity Press. |

| Ref | Reference | Hallucinated Citation |
|---|---|---|
| 48 | Gichangi, R. N., Momanyi, M., & Wambiya, P. (2025). The Extent of Information and Communication Technology (ICT) Integration by Secondary School Teachers in Teaching Practices and Its Influence on Students' Academic Performance." Journal of Popular Education in Africa 9.9 (2025).: 214-230. | Williamson, B., & Piattoeva, N. (2022). Education governance and datafication. Oxford Research Encyclopedia of Education. https://doi.org/10.1093/acrefore/9780190264093.013.1684 |
| 49 | Harvey, S. (2026). Reimagining Education in the Age of AI: Digital Literacy, Decolonization, and the Work Ahead. Atras Journal, 7(1), 294-298. | Williamson, B., & Piattoeva, N. (2022). Education governance and datafication: Critical perspectives on AI and algorithmic systems in schools. Learning, Media and Technology, 47(2), 111–127. |
| 50 | Hasvi, I., Anistia, S., Wicaksana, B. E., & Khotimah, N. (2025, December) Ai And Adaptive Learning: Evaluating Personalized Interventionts On A Distance Education PLATFORM. In Proceeding of the International Conference on Innovation in Open and Distance Learning(Vol. 6, pp. 253-262). | Williamson, B., & Piattoeva, N. (2021). Education governance and datafication. Learning, Media and Technology, 46(3), 219–230. |
| 51 | Hedfeld, P. Artificial Intelligence as a Socio-Economic Dilemma: Ordonomic Diagnosis–Reflection–Design for Education, Work and Governance. Swiss Journal of Business, 80(1):30-49 DOI:10.5771/2944-3741-2026-1-30 | Williamson, B., & Piattoeva, N. (2022). Education governance and datafication. Education and Information Technologies, 27, 3515-3531 |
| 52 | Huang, J., Wang, R. R., Liu, J. H., Xia, B., Huang, Y., Sun, R.,.. & Zou, J. (2025). A Meta-Analysis of LLM Effects on Students across Qualification, Socialisation, and Subjectification. arXiv preprint arXiv:2509.22725 | Williamson B., Piattoeva N. Education governance and datafication. Education and Information Technologies, 27:3515–3531, 2022. |
| 53 | Ivanova, P. (2025). Application Of Generative Ai To Enrich The Educational Process In Out-Of-School Educational Institutions. In Edulearn25 Proceedings (Pp. 4215-4221) Iated | Williamson B., Piattoeva N, Education governance and datafication: The power of data in education. Routledge, 2021 |
| 54 | Jadaun, Rakhi Singh (2025). The Impact of Artificial Intelligence–Powered Learning Tools on Cognitive and Socio-Emotional Development in Children. International Journal Of Innovative Research In Technology Volume 12 Issue 4 ; 151-158 | Williamson, B., & Piattoeva, N. (2022). Education governance and datafication: Critical perspectives on privacy, surveillance, and ethics.Learning, Media and Technology, 47(2), 133–148 |
| 55 | Jerez, S. A. R., Casas, J. S. L., & Osorio, M. R. (2025). Integration of Generative Artificial Intelligence and 3D Immersive Environments in Competency-Based Higher Education. Educational Process: International Journal, 19, e2025544. | Williamson, B. (2021). Education governance and datafication. Routledge. https://doi.org/10.4324/9780429286794 |
| 56 | Kaldi, S. Artificial Intelligent in teaching and learning. In Book of abstracts: Regional ISATT24 Conference (p. 29). | Williamson, B., & Piattoeva, N. (2020). Education Governance and Datafication. Educational Policy, 34(1), 1-28. |
| 57 | Kalonde, Gilbert, Boateng, Samuel & Duedu, Claudia (2025). Artificial Intelligence in Classroom Assessment: Opportunities, Equity Challenges, and Best Practices for Formative and Summative Integration. Open Access Library Journal, 12, 1-16. | Williamson, B. and Piattoeva, N. (2022) Education Governance and Datafication. Education and Information Technologies, 27, 3515-3531. |
| 58 | Kamlanghan, Thanatporn (2025). การนำ AI และ Big Data มา ช่วย ใน การ ตัดสินใจ ของ ผู้ บริหาร สถาน ศึกษา ความ เป็น ไป ได้ และ ข้อ ท้าทาย. ratchasimaparithat, 1(2), 11-21 | Williamson, B., & Piattoeva, N. (2022). Education governance and datafication. Oxford Research Encyclopedia of Education. https://doi.org/10.1093/acrefore/9780190264093.013.1743 |
| 59 | Kasheem, M., Yahya, N., Shalghoum, N., Masuwd, M., Alriteemi, A., Abdullah, M.,.. & Alrumayh, S. (2025). Artificial Intelligence in Academic Research: Adoption, Opportunities, and Barriers among Faculty in Libya Higher Education. Multidisciplinary Journal of Thought and Research, 1(3), 109-127. | Williamson, B., & Piattoeva, N. (2022). Education governance and datafication. Education and Information Technologies, 27, 3515–3531. |
| 60 | Kondo, Y., Dunnell, K., Voysey, I., Hu, Q., Paesano, V., Nguyen, P. H.,.. & Rocher, L. (2026). Interactive visualizations for adolescents to understand and challenge algorithmic profiling in online platforms. arXiv preprint arXiv:2601.07381. | Williamson B., Piattoeva N. 2022. Education governance and datafication. Education and Information Technologies 27 (2022), 3515–3531 |

| Ref | Reference | Hallucinated Citation |
|---|---|---|
| 61 | Kotsini, S. K. S. (2026). Rethinking education: artificial Intelligence, empathy, and emerging ethical challenges." Mediterranean Journal of Education 6.1 (2026).: 32-45. | Williamson, B., & Piattoeva, N. (2022). Education governance and datafication: Artificial intelligence, algorithms, and automation. Educational Philosophy and Theory, 54(3), 221-236. |
| 62 | Kumar, A AI in Overcoming Educational Challenges in Bihar: A Path to Transformative Higher Education." The Chitransh Academic & Research 1(2), 141–161 | Williamson, B., & Piattoeva, N. (2020). Education governance and datafication: Artificial intelligence in education systems. Research in Education, 104(3), 228–245. https://doi.org/10.1177/0034523719876870 |
| 63 | Kumar, M. N., Khan, S. A., & Suresh, N. (2026). Automated Analysis of Student Feedback for Faculty Appraisals Using NLP. In Turning Human Resource Analytics Into Actionable Strategies (pp. 23-56) IGI Global Scientific Publishing | Williamson B. Piattoeva N. (2020). Education governance and datafication.Learning, Media and Technology, 45(1), 1–5. 10.1080/17439884.2020.1707100 |
| 64 | Kumar, S. (2025). Comparative Study Of AI-Based Teaching Methodologies Vs. Traditional Approaches. International Journal of Scientific Research & Engineering Trends, 11(6), 1–6 | Williamson, B., & Piattoeva, N. (2021). Education governance and datafication. In U. Felt, R. Fouché, C. A. Miller, & L. Smith-Doerr (Eds.), The handbook of science and technology studies (pp. 580–606). MIT Press |
| 65 | Kumari, A. (2025). From Guide to Guru? The Dangers of Students Seeing Ai As All-Knowing. Journal Of Advance And Future Research, 3(12), 318-324 | Williamson, B., & Piattoeva, N. (2022). Education governance and datafication: Critical perspectives on AI,automation, and accountability. Learning, Media and Technology, 47(2), 193–206. https://doi.org/10.1080/17439884.2022.2055055 |
| 66 | Lakhiwal, A., Liu, C. W., Bala, H., & Suen, H. Y. (2026). From Opacity to Transparency: Examining the Role of Transparency in Algorithmically Evaluated Asynchronous Video Interviews." Available at SSRN 6088066 (2026). | B Williamson , N Piattoeva. 2022. Education Governance and Datafication. Education and Information Technologies , p. 3515 – 3531 |
| 67 | Lata, K. (2026). AI, Big Data, and Student Privacy: A Policy Perspective in Indian Higher Education. In Ethical, Legal, and Pedagogical Perspectives on AI in Education (pp. 33-62) IGI Global Scientific Publishing | Williamson B. Piattoeva N. (2022). Education Governance and Datafication. Education and Information Technologies, 27, 3515-3531 |
| 68 | Louis, Lavina & Prasad,Druthi (2025). The Impact Of Artificial Intelligence In The Teaching Profession. International Journal of Novel Research and Development, 10(10) b1–b2 | Williamson, B., & Piattoeva, N. (2021). Education Governance and Datafication. |
| 69 | Macedo, F. A., Cruz, M. O. T., & Yana, Y. P. (2025) Apropiación tecnológica y uso académico de las herramientas de inteligencia artificial en estudiantes de Comunicación Puno-Perú. In Entre algoritmos y narrativas: comunicación, identidad y consumo en la sociedad conectada (pp. 26-34). Dykinson | Williamson, B., & Piattoeva, N. (2021). Education governance and datafication. International Journal of Educational Research, 104, 101683. https://doi.org/10.1016/j.ijer.2020.101683 |
| 70 | Maftuhah, L., & Yani, M. (2025). Pilar-Pilar Pendidikan Sepanjang Hayat. PT Arr Rad Pratama | not verified [eBook, not accessible without payment] |
| 71 | Mahalingam, Vasanthi & Bhosale, Rohini (2025). THE ROLE OF ARTIFICIAL INTELLIGENCE IN EDUCATION.Journal Of Emerging Technologies and Innovative Research 12(10),a682-a686 | Williamson, B and Piattoeva, N (2021).Education governance and datafication: The power of data in education. Routledge. |
| 72 | Malagoudar, Aarti F. ; Karadi, Arogyaswamy A. (2025) Transforming Student Learning Through Artificial Intelligence: Evidence from Personalization, Accessibility, and Pedagogical Innovation. International Journal for Multidisciplinary Research, Volume 7, Issue 6, 1–17 | Williamson, B., & Piattoeva, N. (2022). Education governance and datafication: AI, learning analytics, and the rise of algorithmic regulation. Learning, Media and Technology, 47(1), 1–15. https://doi.org/10.1080/17439884.2021.2004556 |
| 73 | Mendonça, A. B. (2026). Environmental Education as a Return to Lifeworld: Nature and Territory. Beyond the Book: How Technology Is Shaping Education, pp. 98–118) Ethics International Press | not verified [eBook, not accessible without payment] |

| Ref | Reference | Hallucinated Citation |
|---|---|---|
| 74 | Mercan, G. (2025). Google VEO 3 As A Pedagogical Partner AI-Driven Storytelling For Conceptual, Affective, And Critical Learning in Science Education. Sinop Üniversitesi Sosyal Bilimler Dergisi, 9(2), 1443-1464. | Williamson, B., & Piattoeva, N. (2021). Education governance and datafication. European Educational Research Journal, 20(4), 435–456. https://doi.org/10.1177/1474904121993792 |
| 75 | Micheal, L. (2025). Student–AI Interaction in the Classroom: Pedagogical Benefits and Concerns." (2025). | Williamson, B., & Piattoeva, N. (2022). Education governance and datafication: Reimagining policy-making and digital technologies. Journal of Education Policy, 37(3), 301–318. |
| 76 | MOLLER-ACUÑA, P. A.; CASTRO VILLARROEL, G. A.; ROJAS-DÍAZ, P. J.; MOTA SUÁREZ, K. T (2024). "Inteligencia artificial en las universidades chilenas: ¿ventaja o desventaja? Nuances: Estudos sobre Educação, Presidente Prudente, v. 35, n. 00, e024019, 2024. e-ISSN: 2236-0441. DOI: https://doi.org/10.32930/nuances.v35i00.10780" | WILLIAMSON, B.; PIATTOEVA, N. Education Governance and Datafication: A Global Perspective. Education Policy, v. 34, n. 3, p. 479–499, 2020. |
| 77 | Moller-Acuña, Patricia & Villarroel, Giordano & Rojas-Díaz, Pedro & Suárez, Katihuska. (2024). Inteligencia Artificial En Las Universidades Chilenas:¿ Ventaja O Desventaja? Inteligência Artificial Nas Universidades Chilenas: Vantagem Ou Desvantagem? Artificial Intelligence In Chilean Universities: Advantage Or." Nuances: Estudos sobre Educação, Presidente Prudente, v. 35, n. 00, e024019, 19 Dec 2024 | WILLIAMSON, B.; PIATTOEVA, N. Education Governance and Datafication: A Global Perspective. Education Policy, v. 34, n. 3, p. 479–499, 2020. |
| 78 | Muhsinin (2025). Digital Humanism In Education: Balancing Artificial Intelligence And Ethical Pedagogy For The 21st Century Classroom. Ogedup: Journal of Global Educational Perspectives 1(1),8–14 | Williamson, B., & Piattoeva, N. (2022). Education governance and datafication: The role of AI in shaping policy and practice. Learning, Media and Technology, 47(4), 439–452. https://doi.org/10.1080/17439884.2022.2040565 |
| 79 | NNAMANI, N. P. (2025). Artificial Intelligence And Future School Management In Secondary Schools In Enugu State." International Nexus Multidisciplinary Research Journal 1.2 (2025).: 109-124. | Williamson, B., & Piattoeva, N. (2022). Education governance and datafication: A comparative analysis of policy technologies in Europe. Journal of Education Policy, 37(2), 194–213. |
| 80 | Obasi, D. H. U. (2023). Ethical Implication Of Artificial Intelligence Implementation in Educational Settings. ISRG Journal of Arts, Humanities and Social Sciences 1(3), 8–14 | Williamson, B., & Piattoeva, N. (2020). Education Governance and Datafication: The Global Emergence of Data-Driven Decision-Making in Education. Global Studies of Childhood, 10(2), 108-120 |
| 81 | Obidovna, D. Z. (2025). Exploring The Dual Impact Of Artificial Intelligence On Educational Sustainability." Shokh Library 1.10 (2025). | Williamson, B., & Piattoeva, N. (2022). Education governance and datafication: AI, accountability, and ethical challenges. New York: Routledge. |
| 82 | ode Riniati, W., & Agusalim, A. (2025). Personalized Learning In The Age Of AI: Opportunities And Challenges For University Students. In International Conference of Business, Education, Health, and Scien-Tech (Vol. 2, No. 1, pp. 1139-1145). | Williamson B., Piattoeva N, Education Governance and Datafication: The Power of Data in Education. London, U.K.: Routledge, 2021. |
| 83 | ÖDEMİŞ KELEŞ, Nurdan (2026). Yükseköğretimde Yapay Zekâ Ve Akademik Dönüşüm. In: Akdağ, Sinan.(Ed) Eğitimde Yapay Zekâ. Ankara. Pp.175–197 | Williamson, B., & Piattoeva, N. (2022). Education governance and datafication. Education and Information Technologies, 27, 3515- 3531 |
| 84 | ODIGILI, B. C., & AKPOTOHWO, F. C. (2025). Reshaping Business Education Undergraduate Students' Future Job Engagement Through Artificial Intelligence. FUO-Journal of Educational Research, 5(2), 180-193. | Williamson, B., & Piattoeva, N. (2021). Education governance and datafication. Policy Futures in Education, 19(1), 5–20. |
| 85 | Panada, R Rupkumar. (2024). Artificial intelligence in educational systems: From early computational tools to contemporary AI-enhanced learning environments. International Journal of Research Publication and Reviews, 5, 3756-3760 | Williamson, B., & Piattoeva, N. (2021). Education governance and datafication: The power of data in education. Routledge. |

| Ref | Reference | Hallucinated Citation |
|---|---|---|
| 86 | Pramanik, S. (2025). Ai, Automation And The Future Of Teaching: Redefining The Teacher's Role." Peace, Values, And Humanity: Reimagining Education For The 21st Century (2025).: 147. | Williamson, B & Piattoeva, N. (2022). Education governance and datafication: Changing data practices and imaginaries. Educational Philosophy and Theory, 54(1), 1-4. |
| 87 | Qahramon, Dinora Nabijonova (2025) Ai Revolution In Teaching. International Journal Of Artificial Intelligence 5(9), 1273-1279 | Williamson, B., & Piattoeva, N. (2022). Education governance and datafication. Learning, Media and Technology, 47(1), 1–6. |
| 88 | Qayyum, A., Sadiqi, T., & Abbas, M. A. (2024). Integrating artificial intelligence into early childhood education policy in Pakistan: Challenges, opportunities, and recommendations." Journal of Development and Social Sciences 5.4 (2024).: 416-431 | Williamson, B., & Piattoeva, N. (2021). Education governance and datafication: Investigating the educational potential and risks of AI. Educational Philosophy and Theory, 53(2), 175–188 |
| 89 | Ramndani, A. E. W., Rizki, J. M., Murtado, M. I. A., Ahyani, R., & Amir, D. R. (2025, November). Dampak Ketergantungan Mahasiswa terhadap Penggunaan AI (Artificial Intelligence) di Bidang Akademik." Seminar Nasional dan Gelar Karya Produk Hasil Pembelajaran. Vol. 3. No. 2. 2025. | Wiliamson, B., & Piattoeva, N. (2023). Education governance and datafication: The new politics of AI and ethics in education. Learning, Media and Technology, 48(1), 1-14. |
| 90 | Rampilla, Y., Satyala, K., & Kailasam, S. (2025). Enhancing Trust in AI-Powered Learning: The Role of Explainable AI in the Education Field." IJSAT-International Journal on Science and Technology 16.3 (2025). | Williamson, B., & Piattoeva, N. (2019). Education Governance and Datafication: Critical Perspectives and Emerging Issues. Learning, Media and Technology, 44(1), 1–14. |
| 91 | Ran,Tao , Basri, Mohd Faizal & Huijun, Li (2025). Exploring the Implications of Artificial Intelligence-GeneratedContent (AIGC) in Educational Practices: Opportunities and Challenges Asian Journal of Research in Education and Social Sciences Vol. 7, No. 7, 388-392, 2025 | Williamson, B., & Piattoeva, N. (2022). Education governance and datafication. In G. W.Noblit (Ed.), Oxford research encyclopedia of education. Oxford University Press. https://doi.org/10.1093/acrefore/9780190264093.013.1545 |
| 92 | Rasheed, Qurrat ul Ain (2025). Stretching Boundaries, Recalibrating Theory: Educational Leadership in the Age of Artificial Intelligence. 1(3), 1–16 AI Research in Educational Leadership https://doi.org/10.5281/zenodo.17541014 | Williamson, B., & Piattoeva, N. (2022). Education governance and datafication. Education and Information Technologies, 27, 3515-3531. |
| 93 | Reddy, K. Future of Global Citizenship Education in a Digitally Divided World." International Journal for Research in Education, Vol. 09, Issue: 11, November 2020 | Williamson, B., & Piattoeva, N. (2020). Education governance and datafication. Oxford Review of Education, 46(4), 463–477 |
| 94 | Roblero, E. G. M., Tenorio, E. J. U., Cevallos, R. P. B., & de la Torre, M. A. C. (2025). La inteligencia artificial (IA) en la investigación educativa dentro del contexto de la educación superior, un estudio documental sistemático. Polo del Conocimiento, 10(5), 1720-1738. | Williamson, B., & Piattoeva, N. (2022). Education governance and datafication: Reimagining the digital future of education. Bloomsbury Publishing. https://doi.org/10.5040/9781350232063 |
| 95 | Rodrigues, Charmain, Mahadi, Rozilawati, Min, Loo Shih (2025). Virtual Assessment Dilemmas in English Literature Writing Courses: A Perspective of the Tertiary level Teachers in Bangladesh. , 9(25), https://doi.org/10.47772/IJRISS.2025.925ILEIID000085 | Williamson, B., & Piattoeva, N. (2022). Education governance and datafication: Shifting roles and ethical challenges of AI-based assessment. Learning, Media and Technology, 47(1), 1–15. https://doi.org/10.1080/17439884.2021.1990133 |
| 96 | Rumbea, G. Z., Ramírez, R. B., Mero, K. Z., & Calero, Y. V. (2025). Aprendizaje autónomo con chatgpt: oportunidades y limitaciones para la educación. Polo del Conocimiento, 10(9), 3346-3360 | Williamson, B., & Piattoeva, N. (2022). Education governance and datafication: The new politics of knowledge. Learning, Media and Technology, 47(1), 1–15 |
| 97 | RUPA, M. K., & KANNADHASAN, M. (2026). Neurodiverse Learners And Ai In English Language Teaching: Towards Inclusive Pedagogies. International Journal of Creative Thoughts 14(1)a915–a923 | Williamson, Ben, and Nelli Piattoeva. "Education Governance and Datafication." Oxford Research Encyclopedia of Education, Oxford UP, 2019 |
| 98 | Rustamovna, A. N. (2025). " Лучшие интеллектуальные исследования 52.2 (2025).: 33-40. | WilliamsonB. Education governance and datafication: The global politics of artificial intelligence. Routledge.2021. |

| Ref | Reference | Hallucinated Citation |
|---|---|---|
| 99 | Saad, Suhana, Shahizan Ali, Mohd Nor, Ramli, Zaimah & Sum, Sarmila Md.2025 A Comparative Analysis Of AI-Powered Adaptive Learning Systems In Higher Education Across Developed Countries. International Journal Of Research And Innovation In Social Science (Ijriss), Volume Ix Issue V, May 2025. | Williamson, B., & Eynon, R. (2020). Education governance and datafication. Learning, Media and Technology, 45(1), 1–14. |
| 100 | Sadiq, A. S., Bello, A. J., & Tukur, S. (2024). Ethical implications of artificial intelligence in the Nigerian education system: A philosophical call for policy integration and curriculum reform. International Journal of Innovative Research in Education, Technology & Social Strategies, 11(1), 171-179. | Williamson, B., & Piattoeva, N. (2020). Education governance and datafication: A critical approach to the politics of knowledge, Educational Philosophy and Theory, 52(4), 350-367 |
| 101 | Sahoo, S., & Behera, K. Artificial Intelligence in Education: Opportunities and Challenges for enhancing teacher competence in Indian classrooms. International Journal of Advanced Academic Studies 2025; 7(10): 118-125 | Williamson B, Piattoeva N. Education governance and datafication. In: Peters M, editor. Encyclopedia of teacher education. Singapore: Springer; 2022, p. 1-5. https://doi.org/10.1007/978-981-13-1179-6_534-1 |
| 102 | Sain, Z. H., Prabowo, J., Hamer, W., Nurlely, L., & Humaera, I. (2025). Integrating ChatGPT in Higher Education: Responsible Use, Ethical Considerations, and Critical Thinking Development. PROCEEDING AISELT, 10(1). | Williamson, B., & Piattoeva, N. (2021). Education governance and datafication. Policy Futures in Education, 19(1), 5–20. |
| 103 | Serantes, M. A. (2025, December). Inteligencia Artificial y su impacto en el trabajo cognitivo desde una perspectiva humanística. In Colloquia, Academic Journal of Culture and Thought (Vol. 12, pp. 309-319). | Williamson, B., & Piattoeva, N. (2023). Education governance and datafication: Critical perspectives. Bloomsbury Academic. |
| 104 | Shabbir, M. S., Naseer, R., Bibi, S., & Faisal, Z. (2025). Towards Responsible Ai In Education: Impacts, Perceptions, And Governance Needs." Contemporary Journal of Social Science Review 3.4 (2025).: 25-39. | Williamson, B., & Piattoeva, N. (2022). Education governance and datafication: The role of AI. Learning, Media and Technology, 47(2), 115–128. https://doi.org/10.1080/17439884.2021.1957595 |
| 105 | Shahata, L. H. A., & Ahmed, E. M. I. (2026). The Role of Artificial Intelligence in Automating and Personalizing Lesson Planning for Teachers. Journal of Faculty of Arts, University of Khartoum, 54(1).199…=310 | Williamson, B., & Piattoeva, N. 2020. Education governance and datafication. Oxford Research Encyclopedia of Education.https://doi.org/10.1093/acrefore/9780190264093.013.470 |
| 106 | Sharma, D. (2026). Climate Change, Migration, And School Disengagement in A Globalised World: Governance Challenges, Educational Inequality, And the Role of AI and Traditional Knowledge. International Journal of Social Sciences, Language and Linguistics, 6(01), 34-43 | Williamson, B. (2021). Education governance and datafication. Routledge |
| 107 | Simos, C., & Katsinis, K. (2026). School Coexistence Without Violence: Relational-Ecological Prevention, Emotional Self-Regulation, and Digitally Mediated Pedagogical Innovation. Open Access Research Journal of Science and Technology, 16(1), 093-106. | Williamson, B., & Piattoeva, N. (2022). Education governance and datafication. Routledge |
| 108 | Sobremesana, Cristopher S.; Padilla, Vanessa B.; Surio, Maricris P. & Abad, Meileen Joy L..(2025) A Framework for Ethical AI in Grade 4 Lesson Planning. Advanced Journal of STEM Education, Vol. 3 No. 1, 70–86. https://doi.org/10.31098/ajosed.v3i1.3428 | Williamson, B., & Piattoeva, N. (2021). Education governance and datafication: A critical perspective on the role of data in educational policy and practice. Routledge |
| 109 | Soomro, A. A., Khan, M. H., Umar, M., Khan, S., & Ali, O. (2025). AI-Driven Academic Advising in Higher Education: Leveraging Intelligent Systems to Personalize Student Support, Improve Retention, and Optimize Career Pathways. The Critical Review of Social Sciences Studies, 3(2), 229-248 | Williamson, B., & Piattoeva, N. (2020). Education governance and datafication. European Educational Research Journal, 19(6), 461–471. |
| 110 | Subrahmanyam, S. (2026). Developing Digital Citizenship Skills. In Advancing Society 5.0 Through AI-Driven Curriculum Innovation (pp. 357-390) IGI Global Scientific Publishing | Williamson B. Piattoeva N. (2020). |

| Ref | Reference | Hallucinated Citation |
|---|---|---|
| 111 | Sultana, S., Memon, A. S., & Bughio, P. A. (2026). Reconceptualizing AI in English Language: Prospects and Challenges in Higher Education. Social Science Review Archives, 4(1), 553-557. | Williamson, B., & Piattoeva, N. (2022). Education governance and datafication. Critical Studies in Education. |
| 112 | Sundaravalli, S. R., Rajasekar, F. V., & Krishna, K. M. (2021). Forthcoming Trends in Education Transformation Using Artificial Intelligence: A Content Analysis." Rigeo 11.12 (2021): 63-77. | Williamson, B., & Piattoeva, N. (2021). Education governance and datafication: Towards a research agenda. Learning, Media and Technology, 46(2), 139-158 |
| 113 | Supriyatin, T., Hidayat, N., Siagian, R. F. E., & Ahmad, D. N. (2026). Sistem Evaluasi Berbasis AI dalam Pendidikan: Dilema Etis dan Implikasi Kebijakan untuk Kepemimpinan Sekolah. JIIP-Jurnal Ilmiah Ilmu Pendidikan, 9(1), 1081-1089. | Williamson B., Piattoeva N. (2022). Education Governance and Datafication. Education and Information Technologies, 27, 3515–3531. https://doi.org/DOI / Link Resmi: http://doi.org/10.1007/s10639-021-10783-6 |
| 114 | Tayyab, I., Ashraf, J., & Aslam, M. (2025). The Interaction Between Self-Regulated Learning Strategies And Ai-Based Learning Assistants: Implications For Metacognitive Development In Education. International Premier Journal of Languages & Literature, 3(2), 816-829. | Williamson, B., & Piattoeva, N. (2022). Education governance and datafication. In R. Lingard, G. Thompson, & S. Sellar (Eds.), Handbook of education policy studies (pp. 1–18). Springer |
| 115 | Temalatzi, R. M., Reyes, A. L., Barranco, A. R., & Xilotl, V. P. (2024). El uso de la inteligencia artificial en los alumnos de nivel superior en la BUAP." AUTE-TIC 1 (2024).: 81-87 | Williamson, B., & Piattoeva, N. (2021). Education governance and datafication: The role of the education data landscape. European Journal of Education, 56(2), 244-257. https://doi.org/10.1111/ejed.12434 |
| 116 | Thakare, P. S., Patil, S., Gorde, S. A., Sudrik, S. R., Jakate, A. K., & Patil, A. S. (2025, February). Edu-Bridge: Personalized Study Recommendations System for Engineering Students." Congress on Control, Robotics, and Mechatronics. Singapore: Springer Nature Singapore, 2025. | Williamson, B., Piattoeva, N.: Education governance and datafication: the global challenge. Eur. Educ. Res. J. 20(2), 153–175 (2021) |
| 117 | Thomas, P. (2025). Fostering Critical Thinking Skills with AI-Augmented Instruction | Williamson, B., & Piattoeva, N. (2020). Education governance and datafication. European Educational Research Journal, 19(6), 677–684 |
| 118 | Thomas, P. (2025). AI in Classroom Management: Intelligent Support for Teacher Efficiency | Williamson, B., & Piattoeva, N. (2022). Education Governance and Datafication. Learning, Media and Technology, 47(1), 1–13 |
| 119 | Tirkeshova, G., Atjanova, A., & Seyran, A. (2025). Artificial Intelligence And The Future Of Teaching And Learning. Наука И Мировоззрение, 1(63), 28-34. | Williamson B., Piattoeva N. Education Governance and Datafication. Routledge, 2021 |
| 120 | Todescu, V. (2025). Predarea Gramaticii Cu Ajutorul Inteligenței Artificiale. Annales Universitatis Apulensis. Series Philologica, 26(3), 328-335. | Williamson B., Piattoeva N, Education Governance and Datafication, în Learning, Media andTechnology, vol. 44, nr. 1, 2019, p. 48 |
| 121 | Turner, M., & David, K. (2025). Educational Governance in a Globalized World: Comparative Perspectives on Accountability, Autonomy, and School Performance. Innovative Journal of Educational Research and Insights, 198-207 | Williamson, B. (2020). Education governance and datafication. Routledge |
| 122 | Valdivia, Pedro José Mayoral (225). "Uso de Inteligencia Artificial en el Análisis de Datos por Estudiantes Universitarios. IN :Incluir, prevenir y transformar. Educación, derechos y justicia social en escenarios de vulnerabilidad> EDS: Rocío Calderón García; María del Carmen Remigio Montero; Jorge Alfredo Jiménez Torres ; Luis Alfonso Pérez Romero 172 –188" | Williamson, B., & Piattoeva, N. (2022). Education governance and datafication: Critical perspectives on data-driven education policy and practice. Routledge |
| 123 | Vattaparambil, Muhammed Sadique Ali (2025). Access to Education: Breaking Barriers with AI. Diversity Research: Journal of Analysis and Trends 3(4) April 2025, 8-11 | Williamson, B., & Piattoeva, N. (2022). Education Governance and Datafication. Bloomsbury. |

| Ref | Reference | Hallucinated Citation |
|---|---|---|
| 124 | Velmurugan, R., Kalimuthu, M., Bhuvaneswari, R., & Abey, J. (2026). Digital Strategies for Supporting Education Aligned With the Sustainable Development Goals (SDGs). In Digital Transformation and Sustainability in Higher Education (pp. 303-332). IGI Global Scientific Publishing | Williamson, B., & Piattoeva, N. (2020). Education governance and datafication: The rise of data- driven decision making in education. Policy Futures in Education, 18(2), 212–230 |
| 125 | Wariza, Janeth M.; Wariza, Robert E. Jr & Salvaleon, Rolly G. (2025) TEACHING WITH AI: LIVED EXPERIENCES, RESPONSIBILITIES, AND CHALLENGES OF TECH-TECHERS. IJARIIE 11(3), 3407–3423 | Williamson, B., & Piattoeva, N. (2021). Education governance and datafication: Rewriting the rulebook for schools. European Educational Research Journal, 20(4), 421–437. |
| 126 | Xalimjon, Abduxamidova Muyassar (2025). LEVERAGING ARTIFICIAL INTELLIGENCE FOR DATA, DECISIONS, AND EFFICIENCY IN SCHOOL GOVERNANCE. RAQAMLI TRANSFORMATSIYA DAVRIDA PEDAGOGIK TA'LIMNI RIVOJLANTIRISH ISTIQBOLLARI. https://doi.org/10.5281/zenodo.17340652 | Williamson, B., & Piattoeva, N. (2022). Education governance and datafication: Reconfiguring accountability and transparency through digital technologies. Policy Futures in Education, 20 (1), 5–20. https://doi.org/10.1177/14782103211065812 |
| 127 | Xu, Y. (2024). Generative Artificial Intelligence empowering inquiry-based learning: Strategy and challenges. Journal of Educational Research and Policies, 6(9), 14-18 | Williamson, B., & Piattoeva, N. Education Governance and Datafication: The Role of Digital Technologies in the Education System. Educational Policy, 2020, 34(5), 949-973 |
| 128 | Yu, X. (2025, September) The Evaluation Dilemma of AI Tools for Teachers: A Critique of Work Simplification and Anti-Design Strategies. In 2025 5th International Conference on Educational Technology (ICET) (pp. 604-608) IEEE. | Williamson B., Piattoeva N "Education governance and datafication: Reimagining education policy through machine intelligence," Crit. Stud. Educ., vol. 63, no. 1, pp. 1–16, 2022 |
| 129 | Abidin, Rohmad, Muhammad Rikzam Kamal, and Abdul Majid. "Aspek Etis Penerapan Kecerdasan Buatan Pada Teknologi Guru Virtual." In UMMagelang Conference Series, pp. 657-662. 2025. | Williamson, B. &. (2021). Education governance and datafication. European Educational Research Journal, 453–467 |
| 130 | Anto, R. P. Bab 5 Dampak Ai Pada Perubahan Paradigma Pendidikan. Artificial Intelligence Dalam Dunia Pendidikan, 87. | |
| 131 | Ayankojo, Fatimah Adewumi & Enu, Evelyn Nwene ( 2025) A Cautious Optimism of Teachers' Perspectives on the Use of AI inLanguage Education. Journal Of International Association Of Language Educators (Jiale), 5(1),pp 10–16 | Williamson, B., & Piattoeva, N. (2022). Education governance and datafication. In M. Apple, W. Au & L. Gandin (Eds), The Routledge International Handbook of Critical Education (pp. 268-280). Routledge |
| 132 | Bazouche, F. (2025). The Socratic Questioning Method in the Age of Artificial Intelligence (AI) in Higher Education. In T. Bastiaens (Ed.), Proceedings of EdMedia + Innovate Learning (pp. 50-58). Barcelona, Spain: Association for the Advancement of Computing in Education (AACE). Retrieved March 24, 2026 from https://www.learntechlib.org/primary/p/226125/ | |
| 133 | Bhadra, M. (2025). From Automation to Innovation: Exploring the Impact of AI. Multidisciplinary Research Area in Arts, Science & Commerce (Volume-6), Biplab Auddya, Raja Priya, Deepti Mujumdar, T Kalpana, Vani S V, Vaishnavi Vitthal Mogal, Shraddha Dnyaneshwar Gawali (eds). Hill Publication pp. 63–65. | Williamson, B., & Piattoeva, N. (2020). Education governance and datafication: The implications of big data and algorithmic decision-making. European Educational Research Journal, 19(4), 379-394. |
| 134 | Cabral, A. L. L., & Peixoto, A. B. (2025). Artificial Intelligence In Academic Work. Cipeex. Resumo Simples | Williamson, B., & Piattoeva, N. (2020). Education governance and datafication: Learning analytics and ethical challenges in higher education. Learning, Media and Technology, 45(3), 211–223. |

| Ref | Reference | Hallucinated Citation |
|---|---|---|
| 135 | Camacho-Navarro, A. (2025). Apropiación de la IA y metodologías docentes: un diagnóstico en Educación Superior. In: Hacia una enseñanza más eficiente, La inteligencia artificial como aliada en la práctica docente Manuel Becerra Polanco, Ricardo Moreno Espinosa & María Leonor Mandujano Rodríguez (eds), Puebla (Mexico): Novelungos Editorial. pp. 67–92 | Williamson, B. (2022). Education governance and datafication: Reimagining learning, surveillance and automation. Critical Studies in Education, 63(1), 1–16. https://doi.org/10.1080/17508487.2020.1835687 |
| 136 | Chandraprakash, K. (2025). The Role Of Self-Concept And Artificial Intelligence In Education. Ai in Education: A Psychological Perspective on Personalized Learning, P. Muthukumar, A. Senthilkumar & K. Malini (eds). Sivakasi: Mithra Dev Publications pp. 378–385 | Williamson, B., & Piattoeva, N. (2022). Education governance and datafication. In L. Hamilton, & A. R. Lund (Eds.), The Palgrave handbook of education and AI (pp. 47–65). Palgrave Macmillan. |
| 137 | Collazos, Leidy Consuelo Torres; Terán, Victoria Eugenia Pino & Lara, Maria Alejandra Alban (2025). Prototipo "Inclusión+" Con Enfoque En Población Lgtbiq+ En La Corporación Universitaria Comfacauca-Unicomfacauca De Popayán- Colombia. In: Pensamiento, cultura y sociedad, Manuel Bermúdez Vázquez & Agustín Sánchez Cotta (eds.).Madrid: Dykinson SL 216–228. | Williamson, B., & Piattoeva, N. (2021). Education governance and datafication. In: Peters, M. A. (Ed.), Encyclopedia of Educational Philosophy and Theory. Springer. |
| 138 | da Piedade Fernandes, Anthony Savio Herminio (2025). Challenges and Opportunities of Implementing AI in Distance Learning. In: AI and Learning Analytics in Distance Learning. Henrique S Mamede & Arnaldo Santos (eds).. IGI Publishing pp. 1–37 | Williamson , B., & Piattoeva, N. (2021 ). Education governance and datafication: The challenges of data privacy. International Journal of Education Policy and Leadership, 16 (1), 1-15. DOI: 10.22230 / ijepl.2021v16n1a404 |
| 139 | de Souza Lemes, S. (2025). Gestão, avaliação ea inteligência artificial: as possibilidades e desafios à mediação pedagógica. Revista Práxis Educacional, 21(52), e18509. 10.22481/praxisedu.v21i52.18509 | WILLIAMSON, Ben; PIATTOEVA, Nelli. Education governance and datafication: critical perspectives. New York: Routledge, 2022 |
| 140 | Digel, S., & Biel, C. (2025). Empowering Learners to Deal with Artificial Intelligence. Adult Education Research on Digitalisation In: Adult Education Research on Digitalisation: Concepts – Scopes – Understandings Franziska Bellinger, Ricarda Bolten-Bühler & Matthias Rohs (eds), WBV pp. 23–44 | Williamson, B. & Piattoeva, N. ( 2021 ). Education Governance and Datafication : Connecting Critical Policy Sociology and Educational Data Studies. British Journal of Sociology of Education, 42 (1), 47-64. |
| 141 | Figueroa, C. A. M. (2025). La inteligencia artificial como estrategia académica en el desarrollo de trabajos de investigación en el nivel de educación superior. In: Hacia una enseñanza más eficiente, La inteligencia artificial como aliada en la práctica docente Manuel Becerra Polanco, Ricardo Moreno Espinosa & María Leonor Mandujano Rodríguez (eds), Puebla (Mexico): Novelungos Editorial. pp., 133–151 | Williamson, B. (2020). Education, Governance and Datafication. In M. A. Peters, T. Besley, & D. Araya (Eds.), The Springer encyclopedia of critical digital pedagogy. Springer. https://doi.org/10.1007/978-981-13-8473-7_92-1 |
| 142 | Goro, G. (2025). The New Literacy in the Age of AI: Educational Foundations, Pedagogical Practices, and Ethical Challenges La nueva alfabetización en tiempos de inteligencia artificial: fundamentos, prácticas y desafíos. New Literacies in the Age of AI: Ethics, Teaching, and Writing (English Version). Rubén González Vallejo (ed.). AG Editor SAS, Montevideo,pp 2–18 | Williamson, B., & Piattoeva, N. (2022). Education governance and datafication in the age of AI. Learning, Media and Technology, 47(4), 365–379. https://bit.ly/3R4r6Jz |
| 143 | Hoca, Fadil & Nuredin, Abdulmecit. (2025). Algorithmic Bias In Ai-Enhanced Education: Cultural Dimensions And Pedagogical Impact. Conference: International Symposium On Law, Justice, And Emerging Global Challenges Symposium Proceedings. International Vision University • Gostivar, North Macedonia • January 21-23, 2025. PP. 163-176 DOI: https://doi.org/10.55843/ISL2025symp163h | Williamson, B., & Piattoeva, N. (2022). Education governance and datafication. European Educational Research Journal, 21(2), 143-160 |

| Ref | Reference | Hallucinated Citation |
|---|---|---|
| 144 | Ramya, M. (2025). Collaboration Between Teachers And Children With Disabilities As A Pathway To Building Inclusive Classrooms And Inclusive Futures. In: Confluence Critical Multidisciplinary Approaches To 21st Century Issues. Jagdish Kumar & Anjaria Mihir Bharatkumar (eds). Red Unicorn Publishing 10.25215/9371832649.23 | Williamson, B., & Piattoeva, N. (2020). Education governance and datafication. Oxford Review of Education, 46(5), 611–627 |
| 145 | Rani, M. N., & Kaur, M. P. (2025) Ai In Personalized Learning And Adaptive Teaching. Shodh Samalochan, 12(1) 22–29 | Williamson, B., & Piattoeva, N. (2021). "Education Governance and Datafication." Routledge. |
| 146 | Singh, S., & Ragi, T. S. (2025) AI-Powered Academic Knowledge Management: Transforming Academic Research and Teaching. In Artificial Intelligence and Human Existence (pp. 33-51). CRC Press | |
| 147 | Семенюк, Р. А. (2025). Етичні питання та ризики використання ШІ в освітньому процесі. Актуальні питання у сучасній науці, (37), 1135-1141. 10.52058/2786-6300-2025-7(37)-1135-1141 | Williamson, B. (2020). Education governance and datafication: knowledge and expertise in the digital age. [in English] |
| A1 | PF, K. A. P. D., Mahardika, I. G. N. A. W., Putra, I. D. G. R. D., Warmayana, I. G. A. K., & Wahyuni, N. N. T. (2025, November) Transformation of Character Education in Elementary Schools: Development of a Model for Strengthening Character and Identity in the Digital Era. In Proceeding of International Conference on Innovation in Elementary Education (Vol. 2, No. 1, pp. 101-109). | Williamson, B., and Piattoeva, N. (2022). Education governance and data infrastructures in the age of ai. Learning, Media and Technology, 47 (1), 1–15. |
| A2 | Qureshi, S., & Irshadullah, H. M. (2025). From AI Ethics to AI Justice: A Comprehensive Framework for Equitable Governance in Education and Social Systems. VFAST Transactions on Education and Social Sciences, 13(3), 55-72. | Williamson, B., and Piattoeva, N. (2022). Education governance and data infrastructures in the age of ai. Learning, Media and Technology, 47 (1), 1–15. |